\newif\ifarxiv
\arxivtrue   
\documentclass[sigconf]{acmart}
\newif\ifarxiv
\arxivtrue    
\AtBeginDocument{%
  }

\copyrightyear{2026}
\acmYear{2026}
\setcopyright{cc}
\setcctype{by}
\acmConference[CHI '26]{Proceedings of the 2026 CHI Conference on Human Factors in Computing Systems}{April 13--17, 2026}{Barcelona, Spain}
\acmBooktitle{Proceedings of the 2026 CHI Conference on Human Factors in Computing Systems (CHI '26), April 13--17, 2026, Barcelona, Spain}
\acmPrice{}
\acmDOI{10.1145/3772318.3791040}
\acmISBN{979-8-4007-2278-3/2026/04}

\usepackage{xcolor}
\usepackage{dblfloatfix}
\usepackage{balance}

\ifarxiv
  \usepackage[disable]{todonotes} 
\else
  \usepackage[colorinlistoftodos]{todonotes}
  \usepackage{fontawesome5}
\fi

\usepackage[utf8]{inputenc} 
\usepackage{amsmath}        

\begin{document}
\title{DuoMorph: Synergistic Integration of FDM Printing and Pneumatic Actuation for Shape-Changing Interfaces}

\author{Xueqing Li}
\orcid{0009-0008-0551-4197}
\email{li-xq23@mails.tsinghua.edu.cn}
\affiliation{%
  \institution{Tsinghua University / The Hong Kong Polytechnic University}
  \city{Beijing / Hong Kong}
  \country{China}
}

\author{Danqi Huang}
\orcid{0009-0005-6583-7359}
\email{hdq24@mails.tsinghua.edu.cn}
\affiliation{%
  \institution{Tsinghua University}
  \city{Beijing}
  \country{China}
}

\author{Tianyu Yu}
\orcid{0000-0003-0609-0994}
\email{tianyuyu@berkeley.edu}
\affiliation{%
  \institution{University of California, Berkeley}
  \city{Berkeley}
  \state{CA}
  \country{USA}
}

\author{Shuzi Yin}
\orcid{0009-0006-4035-0728}
\email{yinsz24@mails.tsinghua.edu.cn}
\affiliation{%
  \institution{Tsinghua University}
  \city{Beijing}
  \country{China}
}

\author{Bingjie Gao}
\orcid{0009-0002-9535-4321}
\email{gbj24@mails.tsinghua.edu.cn}
\affiliation{%
  \institution{Tsinghua University}
  \city{Beijing}
  \country{China}
}

\author{Anna Matsumoto}
\orcid{0009-0008-1348-6368}
\email{antech33@stanford.edu}
\affiliation{%
  \institution{Stanford University}
  \city{Stanford}
  \state{CA}
  \country{USA}
}

\author{Zhihao Yao}
\orcid{0009-0003-9638-7413}
\email{yaozh_h@outlook.com}
\affiliation{%
  \institution{Tsinghua University}
  \city{Beijing}
  \country{China}
}

\author{Yiwei Zhao}
\orcid{0009-0007-8424-3147}
\email{18009238328@163.com}
\affiliation{%
  \institution{Tsinghua University}
  \city{Beijing}
  \country{China}
}

\author{Shiqing Lyu}
\orcid{0009-0002-6916-5789}
\email{lvsq22@mails.tsinghua.edu.cn}
\affiliation{%
  \institution{Tsinghua University}
  \city{Beijing}
  \country{China}
}

\author{Yuchen Tian}
\orcid{0009-0009-7695-5675}
\email{1849859198@qq.com}
\affiliation{%
  \institution{Tsinghua University}
  \city{Beijing}
  \country{China}
}

\author{Lining Yao}
\orcid{0000-0002-8842-2317}
\email{liningy@berkeley.edu}
\affiliation{%
  \institution{University of California, Berkeley}
  \city{Berkeley}
  \state{CA}
  \country{USA}
}

\author{Haipeng Mi}
\orcid{0000-0003-0560-4228}
\email{haipeng.mi@gmail.com}
\affiliation{%
  \institution{Tsinghua University}
  \city{Beijing}
  \country{China}
}

\author{Qiuyu Lu}
\orcid{0000-0002-8499-3091}
\email{qiuyu.lu@polyu.edu.hk}
\affiliation{%
  \department{Interbeing Lab, School of Design}
  \institution{The Hong Kong Polytechnic University}
  \city{Hong Kong}
  \country{China}
}

\ifarxiv
  \authornote{Corresponding author}
\else
  \authornote{\faIcon[regular]{envelope}\ Corresponding author}
\fi


 \renewcommand{\shortauthors}{Li et al.}


\begin{abstract}
We introduce DuoMorph, a design and fabrication method that synergistically integrates Fused Deposition Modeling(FDM) printing and pneumatic actuation to create novel shape-changing interfaces. In DuoMorph, the printed structures and heat-sealed pneumatic elements are mutually designed to actuate and constrain each other, enabling functions that are difficult for either component to achieve in isolation. Moreover, the entire hybrid structure can be fabricated through a single, seamless process using only a standard FDM printer—including both heat-sealing and 3D/4D printing.
In this paper, we define a design space including four primitive categories that capture the fundamental ways in which printed and pneumatic components can interact. To support this process, we present a fabrication method and an accompanying design tool. Finally, we demonstrate the potential of DuoMorph through example applications and performance demonstrations. 

\end{abstract}


\begin{teaserfigure}
\includegraphics[width=\textwidth]{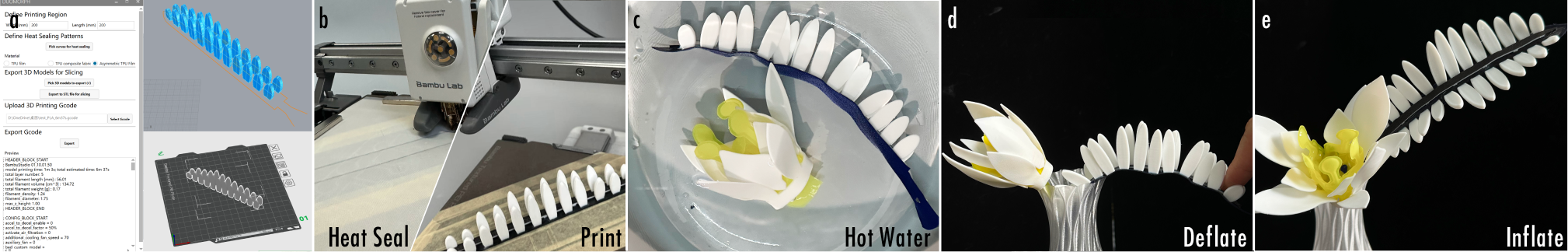}
  \caption{DuoMorph overview. (a) Design tool for modeling and generating integrated G-code; (b) Fabrication of heat-sealed pneumatic actuators (left) and subsequent printing of structures directly on top (right) using an FDM printer; (c) Preshaping the pneumatic actuator with 4D-printed structures in hot water; (d, e) A mimosa-inspired interactive sculpture that folds when touched and blooms when left alone. The flower is also fabricated through a same process.}
  \Description{DuoMorph overview.A sequence of images showing the DuoMorph fabrication process. (a) Screenshot of a software interface for generating G-code and visualizing models. (b) A fabrication setup with heat-sealed pneumatic actuators and an FDM printer adding structures on top. (c) The actuator being heated in water for 4D shaping. (d–e) A flower-like pneumatic sculpture that folds when deflated and blooms when inflated.}
   \label{fig:teaser}
\end{teaserfigure}

\maketitle

\section{Introduction}
Pneumatic interfaces have gained significant traction in Human-Computer Interaction (HCI) and soft robotics due to their lightweight, compliance, and reversible actuation~\cite{yao_pneui_2013, yang_snapinflatables_2024, ou_aeromorph_2016}. They have enabled a wide range of applications, from shape-changing displays~\cite{yao_pneui_2013, gu_pneumesh_2022, lu_millimorph_2019, milliware} and haptic feedback~\cite{delazio_force_2018, he_pneuhaptic_2015, choi_aspire_2021} to multifunctional robotic systems~\cite{lu_sustainflatable_2023, lu_fluidic_2023, lu_millecrepe_2024}. However, fabricating soft pneumatic actuators remains challenging. Casting approaches support complex geometries but require molds and labor-intensive processes~\cite{yao_pneui_2013, moradi_siloseam_2020}. 3D printing provides digital fabrication capabilities, yet reversible pneumatic actuators often rely on costly elastic-resin-based processes such as SLA (Stereolithography)~\cite{vazquez_3d_2015}, while FDM-printed actuators are typically irreversible due to material constraints\cite{wang_pneufab_2023}. In the meanwhile, heat sealing offers a low-cost and accessible alternative that can produce diverse actuation behaviors with simple tools, and is often adopted in making shape-changing interfaces \cite{lu_millimorph_2019, ou_aeromorph_2016, niiyama_sticky_2015, yamaoka_accordionfab_2018}.

Recent efforts have adapted consumer-grade FDM printers for heat sealing, lowering barriers to entry~\cite{choi_therms-up_2021, yang_snapinflatables_2024}. However, these works largely stop at heat sealing alone, without exploring the integration of heat sealing with FDM printing.  This raises several design curiosities: What happens if we continuously print on sealed air bags? How should the structural design and fabrication process be tuned? How reliably can printed structures bond with the airbags? And most importantly, what new capabilities emerge when the two are combined?

Motivated by these curiosities, we developed DuoMorph, a design and fabrication method that synergistically integrates FDM printing and pneumatic actuation to create novel shape-changing interfaces (Fig.~\ref{fig:teaser}). In DuoMorph, printed structures and pneumatic elements are designed to actuate and constrain each other, enabling functions that neither could hardly achieve alone. Moreover, the entire hybrid structure can be fabricated in a single, seamless process using only a standard FDM printer—including both heat sealing and 3D/4D printing. \textcolor{black}{This approach enables a streamlined, more precise, and more consistent process, while significantly reducing labor compared to preparing pneumatic actuators and printed parts separately and then manually gluing and assembling them.}

Based on this principle, we define a design space comprising four primitive categories that capture the fundamental ways in which pneumatic and printed components can interact. To support this process, we present an integrated fabrication method and a design tool that automatically generates G-code for production. Finally, we demonstrate DuoMorph’s potential through a series of application examples and performance evaluations.

The contributions of this work are threefold:
\begin{itemize}
    \item A synergistic design strategy that leverages the interaction between pneumatic actuators and FDM printed structures to enable functions that either component alone can hardly achieve.
    \item An integrated fabrication workflow and accompanying design tool that unify heat sealing and FDM printing into a single streamlined process.
    \item Application demonstrations and evaluations that showcase the expressive and functional capabilities of DuoMorph in interactive systems.

\end{itemize}

\section{Related Works}
\textbf{Pneumatic Interfaces.} Pneumatic interfaces, emerging from the field of soft robotics, have been widely studied in HCI \cite{ma_computational_2017, teng_pupop_2018, luo_digital_2022, follmer_jamming_2012}, due to their potential to safely bridge the gap between machines and people, a capability rooted in their inherent compliance and adaptability \cite{rus_design_2015}. Inflatable structures particularly offer several advantages for interaction design, including being lightweight, cost-effective, scalable, and easily stored \cite{yang_snapinflatables_2024, baines_rapidly_2021}. However, the fabrication and control of these interfaces have been central research challenges that have evolved over time \cite{rus_design_2015}.

Fabrication of early pneumatic actuators often involved techniques such as molding and casting of silicone rubbers, which required intensive manual labor and limited design flexibility for complex geometries \cite{rus_design_2015, xavier_3d-printed_2021}. Meanwhile, researchers explored a more accessible and rapid approach using heat-sealed thermoplastic films with embedded programmable folds, which allows the fabrication of inflatable geometries \cite{ou_aeromorph_2016, sareen_printflatables_2017, yao_pneui_2013, yang_snapinflatables_2024}. This approach was extended with multilayer techniques that produce more complex 3D deformations and functionalities \cite{zhao_multipneu_2022, lu_millecrepe_2024}, and was further developed by repurposing standard FDM 3D printers as heat-sealing tools \cite{ ozdemir_speed_2024, choi_therms-up_2021}.
 
While much of the research focused on the geometry of the seal itself, some research has also explored how to enhance the capability of these pneumatic structures by integrating external or additional components. One approach involved the addition of non-inflatable materials to control deformation. For example, PneUI added multilayer structures with different mechanical or electrical properties, such as paper or conductive thread \cite{yao_pneui_2013}, while BlowFab engraved patterns with a resistant resin or heat-resistant film onto plastic sheets to control the bending and texture of the final shape \cite{yamaoka_blowfab_2017}. Others have added separate mechanical components to enable more complex behaviors, such as passive check valves for sequential control \cite{chen_pneuseries_2021}. Furthermore, fluidic logic systems enabled electronics-free control through custom valves and circuits \cite{decker_programmable_2022, lu_fluidic_2023, chen_pneuseries_2021}. Specific approaches include miniaturizing fluidic chambers for high-resolution, sequential control \cite{lu_millimorph_2019}, and more recently, designing pneumatic "coding blocks" that function like software 'If' or 'For' statements, enabling autonomous, decision-making behaviors \cite{picella_pneumatic_2024}.

To support creativity and rapid prototyping, previous research, including Pneuduino, pioneered in modular hardware toolkits to manage complex systems for pneumatic interfaces \cite{ou_tei_2016, holland_soft_2017, kopic_inflatibits_2016}, and further research was done to make them compact and more comprehensive \cite{tian_openpneu_2023, morita_inflatablemod_2023, shtarbanov_flowio_2021}. PneuBots \cite{youn_pneubots_2022} introduced a construction kit with various inflatable modules and pneumatic connectors and splitters, allowing users to assemble diverse structures.

\textbf{FDM/SLA Printed Interactive Devices.} Separate from inflatables, FDM printing has been widely explored as a low-cost and accessible approach for fabricating interactive devices. A significant research direction explores 4D printing, where printed objects are programmed to transform over time in response to external stimuli \cite{momeni_review_2017}. FDM has been essential in this domain for creating heat-activated composites, enabling complex deformations such as the self-folding from 2D to 3D \cite{an_thermorph_2018}, the programmable bending of linear structures \cite{wang_-line_2019}, and the morphing of mesh surfaces \cite{wang_4dmesh_2018}. Beyond shape transformation, the FDM process offers fine control, as seen in quasi-woven textiles \cite{forman_defextiles_2020}, 3D weaving \cite{takahashi_3d_2019}, tunable hair-like structures \cite{wang_x-hair_2024}, and expressive textures created by manipulating extruder parameters \cite{takahashi_expressive_2017}.

More recently, research has been focused on embedding pneumatic functions directly into FDM-printed artifacts. This includes systems that integrate electronics-free pneumatic logic gates and I/O directly into a 3D model's geometry \cite{savage_airlogic_2022} and toolkits that provide libraries of printable pneumatic widgets for rapid prototyping \cite{kim_morpheesplug_2021}. While these systems successfully embed pneumatic logic or actuation, they remain distinct from methods that create inflatable structures from soft films, requiring the combination of the FDM print with a separate process like blow molding \cite{wang_pneufab_2023}. Beyond FDM, Stereolithography (SLA) has also been utilized to create fluid-driven devices \cite{yan_fabhydro_2021}.

\textbf{Combination of FDM/SLA printing and Pneumatic Actuation.} A smaller body of work has sought to combine 3D-printing and pneumatic actuation explicitly. Stereolithography (SLA) has been used to fabricate reversible pneumatic actuators directly, though these methods often rely on specialized and costly elastic resins \cite{yan_fabhydro_2021}. FDM-based approaches, in contrast, \textcolor{black}{while offering the potential for multi-material printing}, have \textcolor{black}{still} typically produced actuators that are either part of a multi-step, hybrid workflow—such as creating pre-forms for blow molding—or whose shape changes are irreversible due to material constraints \cite{wang_pneufab_2023}. Other methods repurpose the printer as a single-function tool, using its heated nozzle solely to heat-seal separate thermoplastic films \cite{yang_snapinflatables_2024, choi_therms-up_2021} \textcolor{black}{ or a speed‑modulated ironing tool \cite{ozdemir_speed_2024}}. Therefore, the challenge of a single, unified, low-cost workflow that uses a standard FDM printer to seamlessly fabricate 3D structural components and create reversible, heat-sealed inflatables remains largely unexplored. To address this gap, we introduce a novel methodology that seamlessly integrates FDM printing and heat-sealing in a single, accessible process.
\section{Design Space}
\begin{figure}[!htb]
    \centering
    \includegraphics[width=1\linewidth]{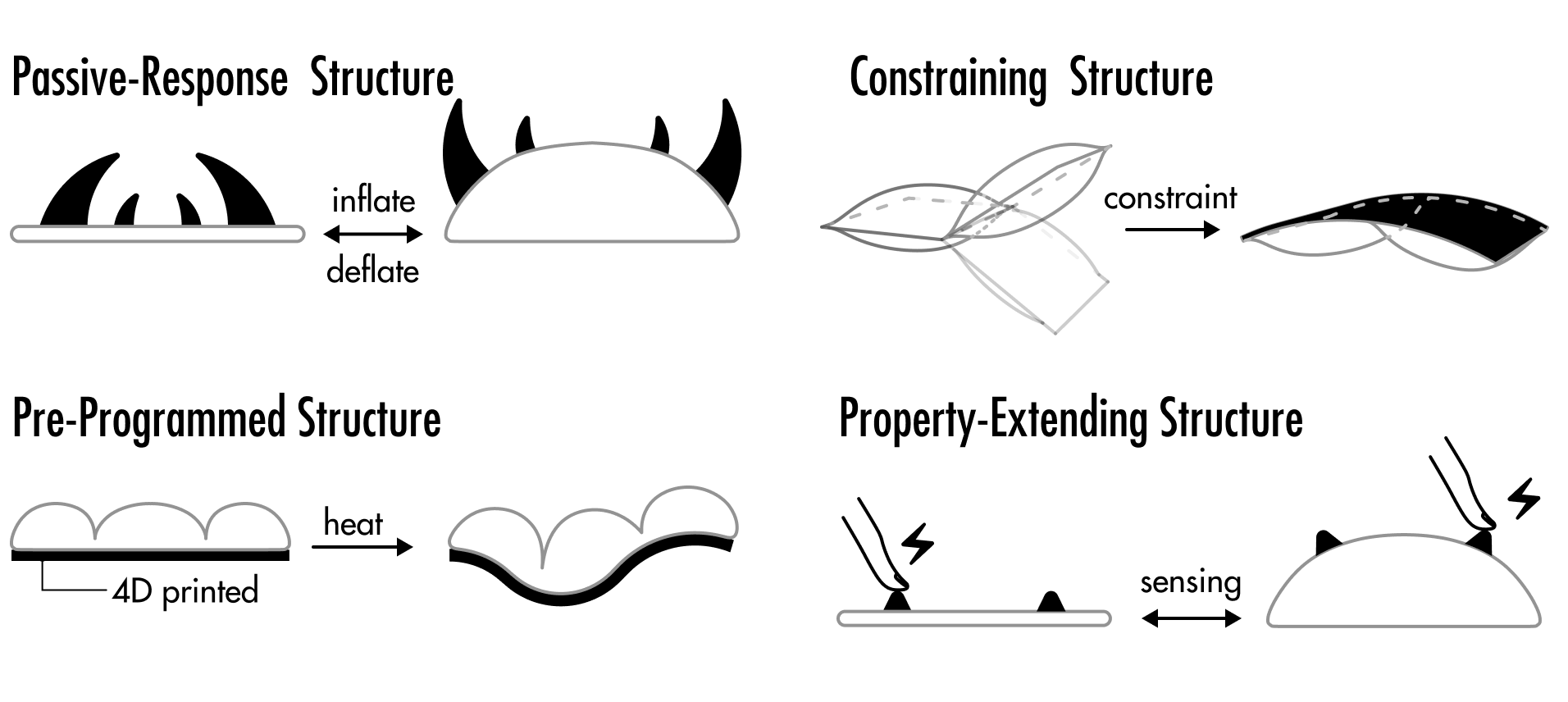}
    \caption{\textcolor{black}{The four-dimensional design space of DuoMorph, categorized based on how the printed structures (black) would interact with the pneumatic actuators (gray)} .}
    \Description{Design space of DuoMorph.Diagram illustrating four categories of printed structures interacting with pneumatic actuators: (top left) passive deformation structures that move with inflation and deflation, (top right) constraining structures that guide or restrict motion, (bottom left) active deformation structures with 4D-printed responses to heat, and (bottom right) function-extending structures responding to external stimuli.}
    \label{fig:design_space_overview}
\end{figure}

Based on how the printed structures interact with the pneumatic actuators, we propose a four-dimensional design space, categorized according to the functions served by the printed structures
\textcolor{black}{as shown in} Fig.~\ref{fig:design_space_overview}:
\begin{itemize}

  \item\textcolor{black}{Passive Deformation Structures}, which respond to deformation of the pneumatic actuator;
    
  \item \textcolor{black}{Constraining Structures}, which regulate deformation of the pneumatic actuator;
  
  \item\textcolor{black}{Pre-shaping Structures}, which impose a preset geometry before actuation;
  
  \item \textcolor{black}{Function-Extending Structures}, which extend the capabilities of the device beyond deformation
  
\end{itemize}
This section will explain these dimensions with example primitives.

\subsection{Passive Deformation Structure}
\begin{figure}[!htbp]
    \centering
    \includegraphics[width=1\linewidth]{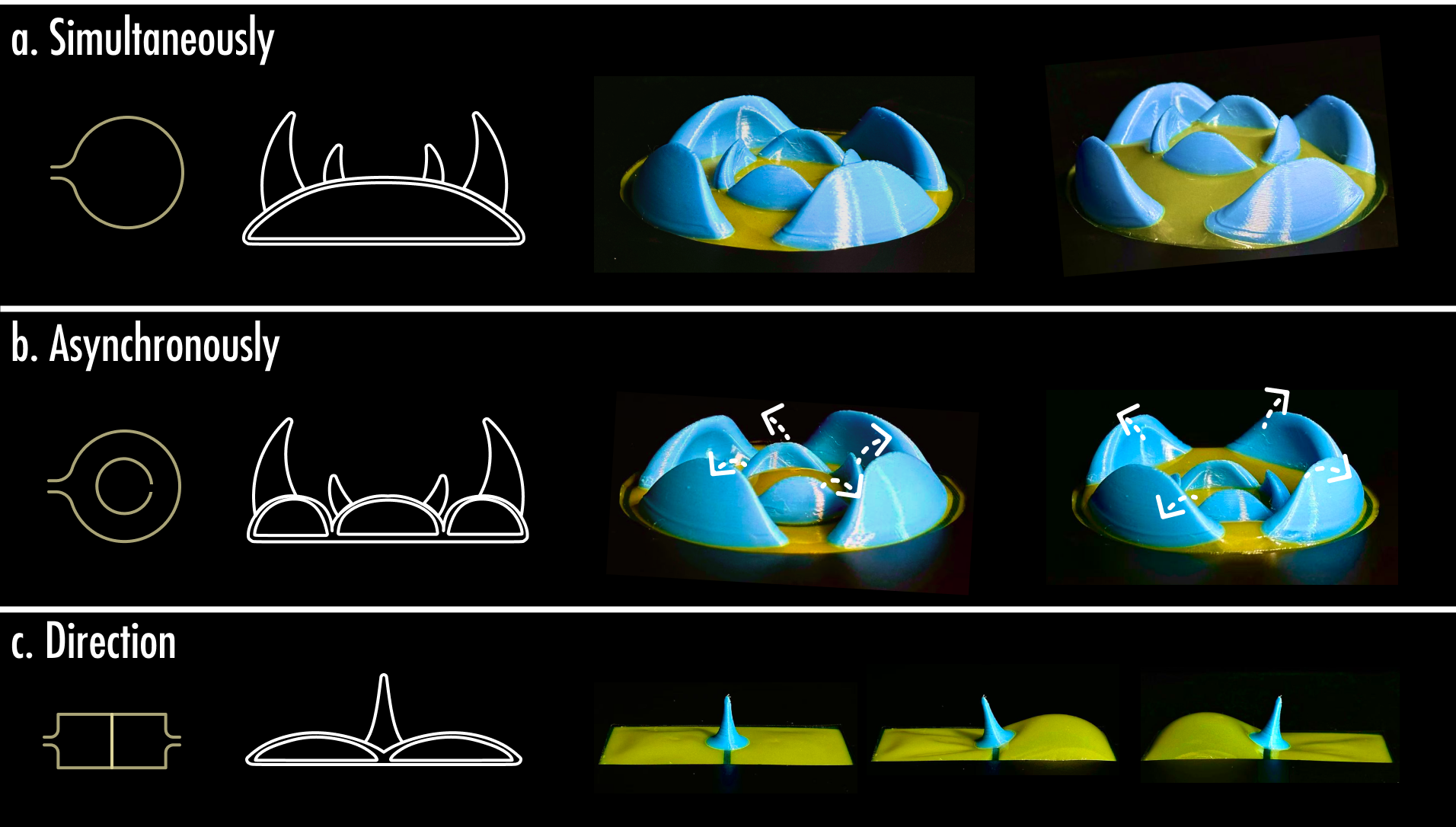}
    \caption{\textcolor{black}{Example printed passive structure. The structures are actuated (a) simultaneously, (b) asynchronously, and (c) toward different directions} by the airbags.}
    \Description{Passive deformation examples.Three rows of images showing airbag-actuated printed structures. (a) Structures deform simultaneously, (b) deform asynchronously, and (c) bend in different directions. Diagrams accompany each case, showing airbag layouts and deformation patterns, followed by photos of blue and yellow prototypes in motion.}
    \label{fig:ds_passive}
\end{figure}

The printed structure can locomote or deform as the pneumatic actuator inflates or deflates. For example, Fig.~\ref{fig:ds_passive}.a shows the printed petals blooming simultaneously when the airbag is inflated. If the structures are distributed across different chambers, they can also be actuated asynchronously (Fig.~\ref{fig:ds_passive}.b). Moreover, a single structure may span multiple chambers; in this case, it can move in different directions depending on which chamber is inflated (Fig.~\ref{fig:ds_passive}.c).

\subsection{Constraining Structure}
\begin{figure}[t]
    \centering
    \includegraphics[width=1\linewidth]{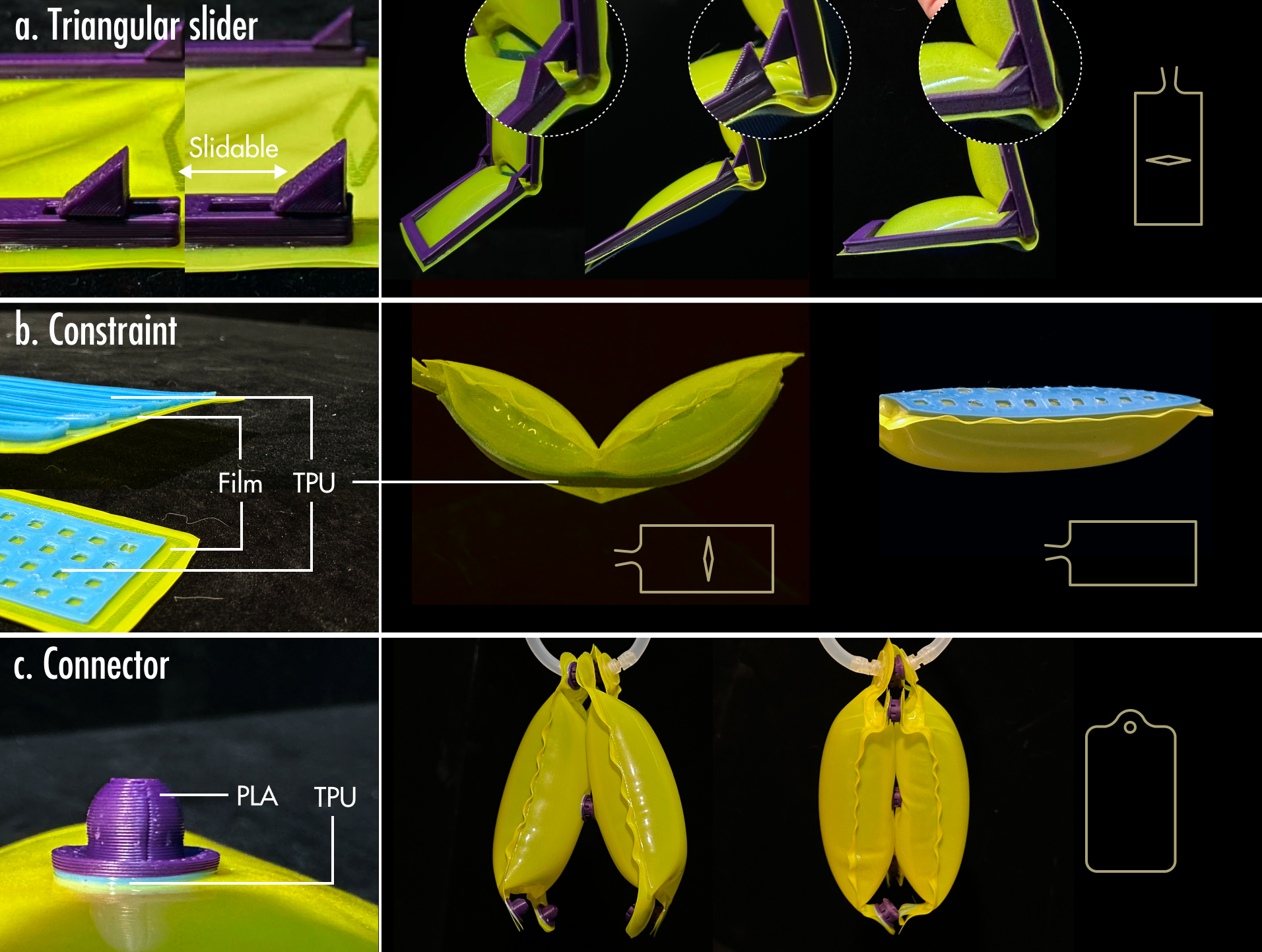}
    \caption{\textcolor{black}{Example printed constraint structures that can control how the airbags deform when inflated include: (a) structures that allow tuning of the bending angle, (b) patterns that change or reverse the bending direction, and (c) connectors that enable assembling multiple airbags into more complex configurations.}}
    \Description{Constraint structure examples.Images demonstrating how printed structures control airbag deformation. (a) Triangular sliders adjust bending angles, (b) perforated TPU film layers constrain bending direction, and (c) connectors link multiple airbags together. Each row shows annotated photos and schematic insets of the structures.}
    \label{fig:ds_constraint}
\end{figure}

The printed structures can also influence how pneumatic actuators deform when inflated, serving as constraints. First, they can be used to reconfigure the bending angle of an airbag, which is conventionally tuned through sealing patterns and becomes fixed once sealed. As shown in Fig. ~\ref{fig:ds_constraint}.a, a triangular slider along a rail can stop the airbag at different angles, offering tunability. Second, continuous printed layers on the airbag surface can reinforce specific regions, thereby controlling the actuator’s bending direction (Fig.~\ref{fig:ds_constraint}.b). Finally, connectors can be directly printed onto the airbag, enabling easy assembly of multiple airbags into various configurations (Fig.~\ref{fig:ds_constraint}.c).

\subsection{Pre-shaping Structure}
The pre-shaping deformation structure can deform the pneumatic actuator into desired shapes using 4D printing techniques. \textcolor{black}{Although pneumatic actuators can achieve 3D deformation independently, integrating 4D-printed structures offloads part of this responsibility, simplifies actuator design, and frees internal chambers to focus on other functions (e.g., enhancing the movement of passive-deformation structures)}. 

\begin{figure}[t]
    \centering
    \includegraphics[width=1\linewidth]{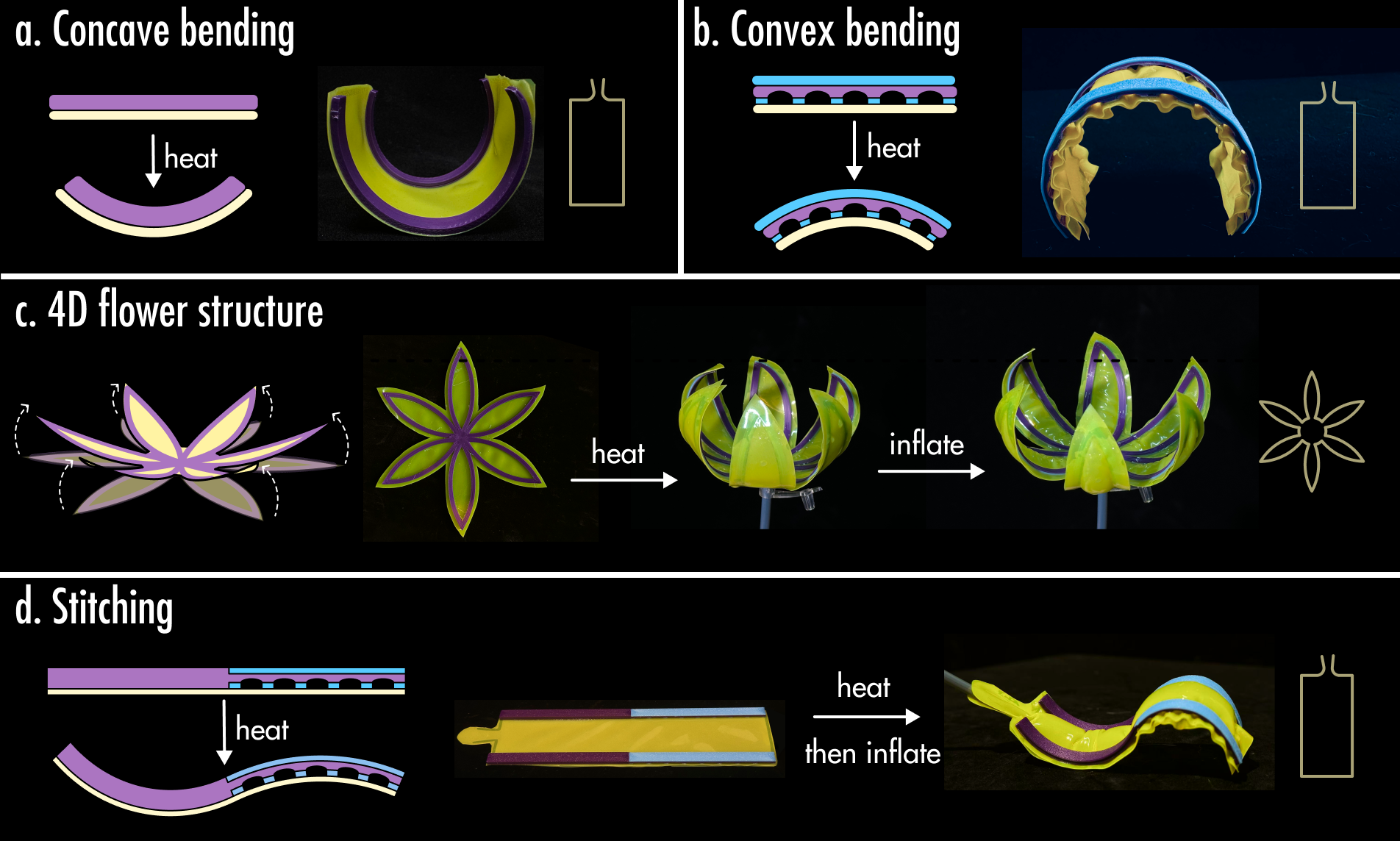}
    \caption{\textcolor{black}{Example 4D printed pre-shaping structures that can preshape the airbags. When heated, the primitives (a) Bend toward the printed structure side (concave bending); (b) Bend toward the other side (convex bending); (c,d) Deform to complex shapes}.}
    \Description{Active 4D structures.Illustrations and photos of 4D-printed structures that change shape when heated. (a) Concave bending toward the printed side, (b) convex bending away, (c) a 4D flower that opens upon heating and inflation, and (d) stitched structures that curve and deform into complex shapes after heat activation.}
    \label{fig:ds_active}
\end{figure}

Fig~\ref{fig:ds_active} shows some example designs. In Fig.~\ref{fig:ds_active}.a, the actuator made from a TPU film effectively serves as—and replaces—the TPU layer in conventional 4D printed structures. When heated, the printed PLA layer contracts while the TPU film remains stable, forcing the pneumatic actuator to bend upward to the side with the printed structures (concave bending). The achievement of bending in the opposite direction (convex bending) is less straightforward. In conventional 4D printing, this can be done by simply switching the TPU layer from the bottom to the top. However, in the DuoMorph fabrication process, the TPU film can serve only as the bottom layer. To address this limitation, we propose the structure shown in Fig.~\ref{fig:ds_active}.b: here, the bottom continuous-arched pattern enables PLA to adhere to the TPU film discontinuously, allowing this side to contract upon heating. Meanwhile, the top side has a printed TPU layer that constrains PLA shrinkage, causing the actuator to bend downward. By combining these basic deformation modes, more complex shape changes can be achieved (Fig.~\ref{fig:ds_active}.c, d).

\subsection{Function-Extending Structure}
\begin{figure}[t]
    \centering
    \includegraphics[width=1\linewidth]{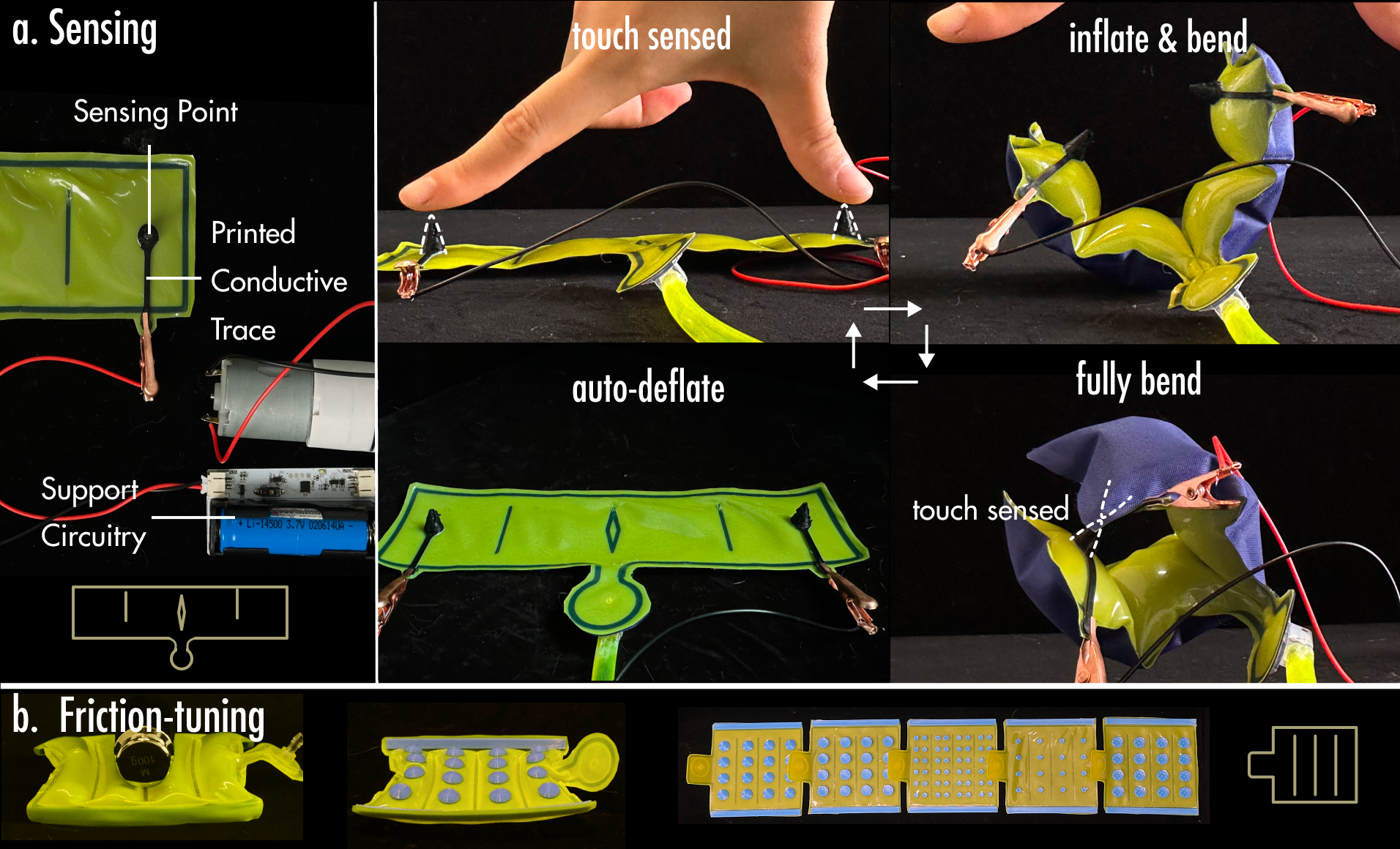}
    \caption{\textcolor{black}{Example printed enabling structures that extend the capabilities of pneumatic actuators beyond deformation: (a) adding sensing capability, (b) tuning surface friction}.}
    \Description{Function-Extending Structure. A diagram showing two examples of printed structures that enhance pneumatic actuators. Part (a) illustrates a green actuator with printed sensors that detect touch, causing it to inflate, bend, and then auto-deflate. Part (b) displays a yellow actuator with different surface textures to tune friction.}
    \label{fig:ds_enabling}
\end{figure}

The printed structures can be leveraged to extend the capabilities of pneumatic actuators beyond deformation, enabling new functions. For example, by using conductive filament, structural features and circuit traces can be directly printed onto the airbag to add sensing capabilities. As shown in Fig.~\ref{fig:ds_enabling}.a, one primitive demonstrates an airbag that inflates to bend when touched, then automatically deflates and returns after fully bending. Another direction we explored is tuning the surface friction of the airbag. Fig. \ref{fig:ds_enabling}.b illustrates a series of primitives with dot arrays of varying sizes and densities. Experiments show that different patterns produce distinct levels of friction adjustment (section.\ref{sec:friction_test})
\section{Fabrication}

\subsection{Walkthrough}

\begin{figure}[b]
    \centering
    \includegraphics[width=1\linewidth]{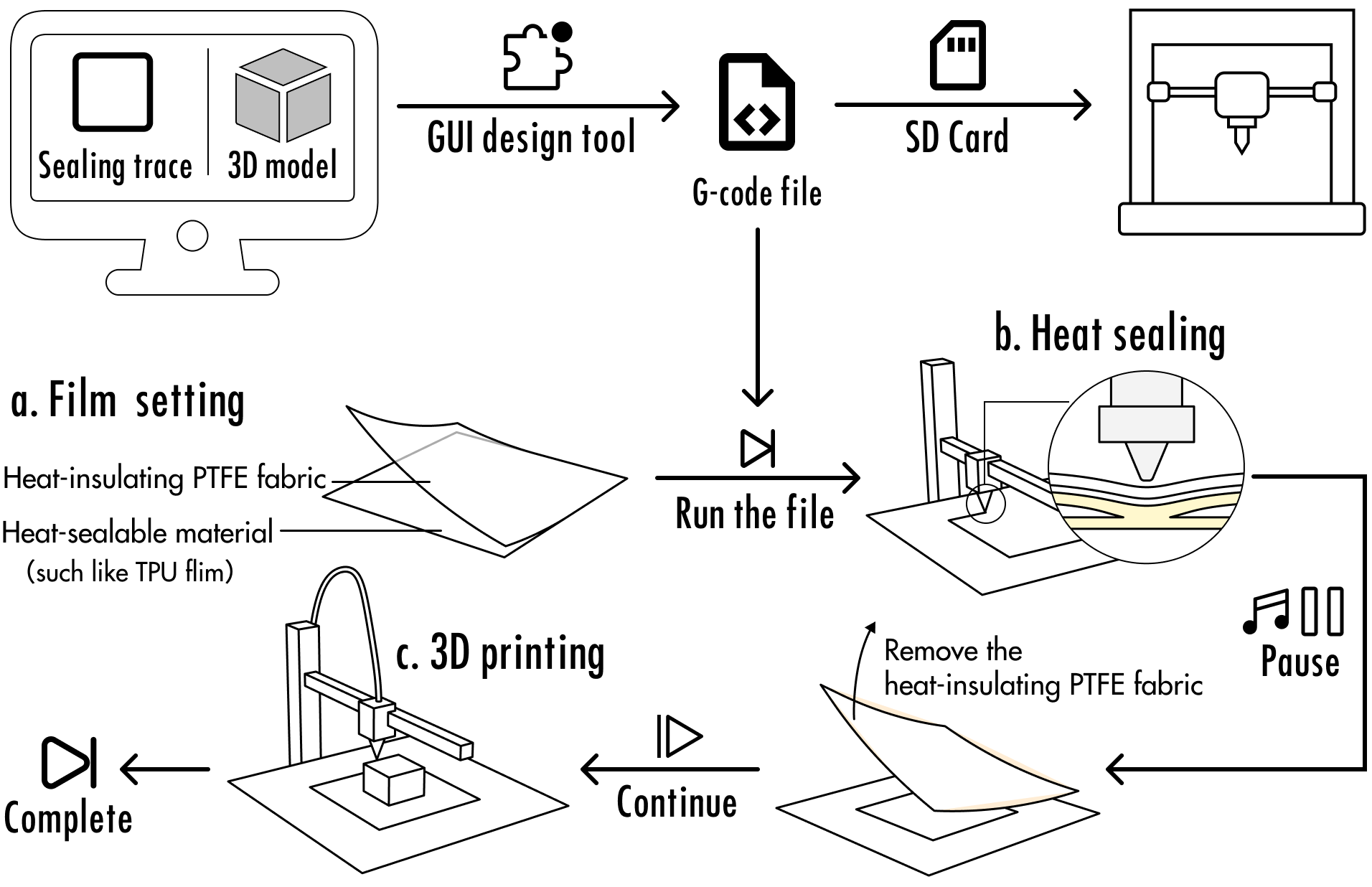}
    \caption{The overall fabrication pipeline.}
    \Description{A flowchart illustrating the fabrication pipeline. It begins with a GUI design tool creating a G-code file from a sealing trace and 3D model. The file is sent to a 3D printer via SD card. The process includes setting heat-insulating PTFE fabric and heat-sealable material, running the file for heat sealing, removing the PTFE fabric, and continuing to 3D print the final part.}
    \label{fig:fab_walkthrough}
\end{figure}

Our overall workflow is illustrated in Fig.~\ref{fig:fab_walkthrough}, achieving \textbf{integrated fabrication of heat sealing and printing} through FDM printers. In the design phase, we developed and integrated multiple parametric tools (such as heat sealing and 4D printing) in the Rhino and Grasshopper environment, and built a user interface based on HumanUI to support integrated operations and improve production efficiency. \textcolor{black}{During fabrication, we used a standard single-nozzle Bambu Lab A1 Series 3D printer for integrated heat sealing and printing. If multiple filament types are required, manual filament swapping or Bambu Lab’s AMS (Automatic Material System) can be used. The musical prompt function acts as an audio reminder for the next step so the user doesn’t have to keep an eye on the printer}.

\subsection{Digital Design}

\begin{figure}[b]
    \centering
    \includegraphics[width=1\linewidth]{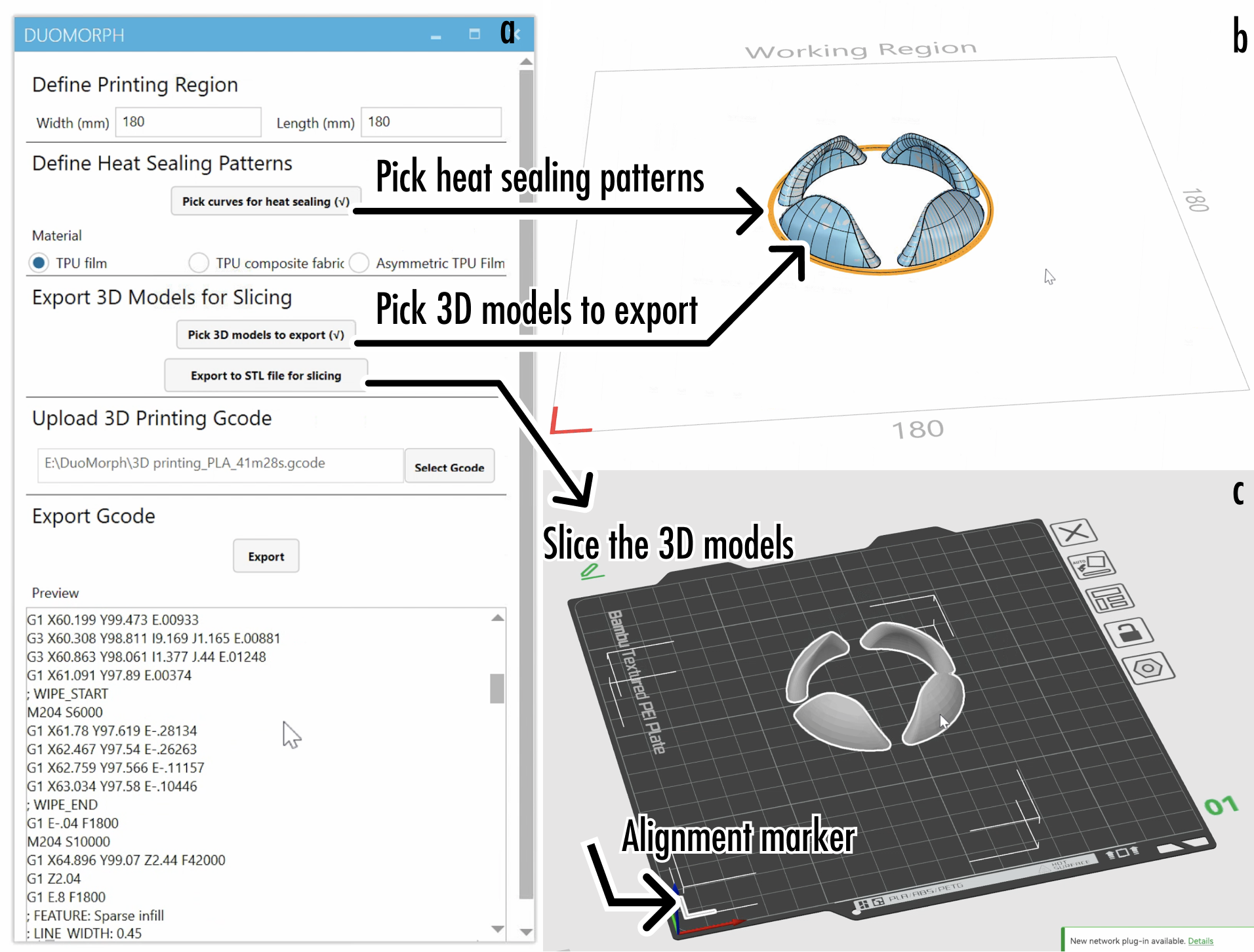}
    \caption{The design tool. (a) DuoMorph design tool GUI, which automatically (b) generates the G-code from sketches to fabricate the heat sealing patterns and (c) merges with the G-code for the 3D printing parts generated by general slicing software.}
    \Description{Screenshots of the DuoMorph design tool. Panel (a) shows the GUI where users define printing regions, select materials, and export G-code. Panel (b) displays a 3D model with highlighted curves for heat sealing. Panel (c) shows slicing software merging the generated G-code with 3D printing instructions, including an alignment marker on the build plate.}
    \label{fig:design tool}
\end{figure}

\begingroup
\setcounter{figure}{9} 
\begin{figure*}[b]
    \centering
    \includegraphics[width=1\linewidth]{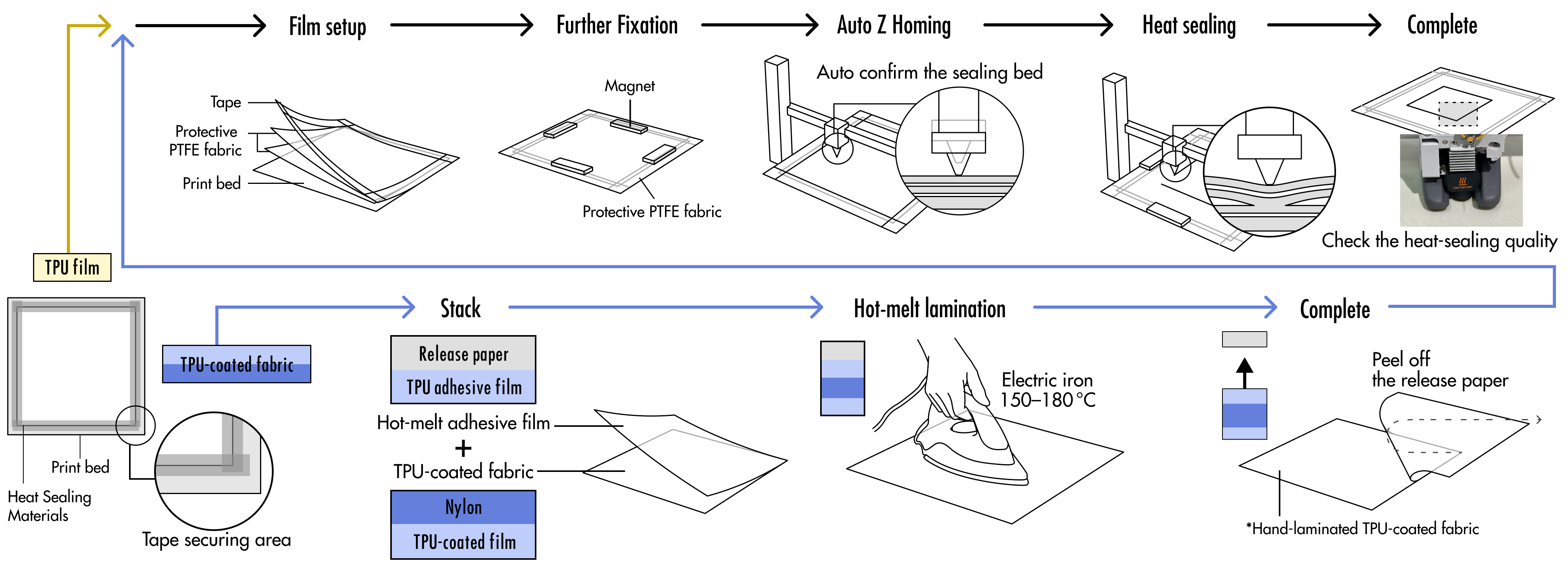}
    \caption{The detailed workflow of heat-sealing.}
    \Description{A detailed workflow diagram for heat-sealing. It outlines steps: film setting with tape and PTFE fabric, Z-offset adjustment, magnet fixation, heat sealing, and completion. An alternative path shows stacking materials (release paper, adhesive film, nylon, coated fabric), applying hot-melt lamination with an electric iron, peeling off release paper, and completing hand-laminated TPU-coated fabric.}
    \label{fig:fab_film_setting}
\end{figure*}
\endgroup

\begingroup
\setcounter{figure}{8}
\begin{figure}[t]
    \centering
    \includegraphics[width=1\linewidth]{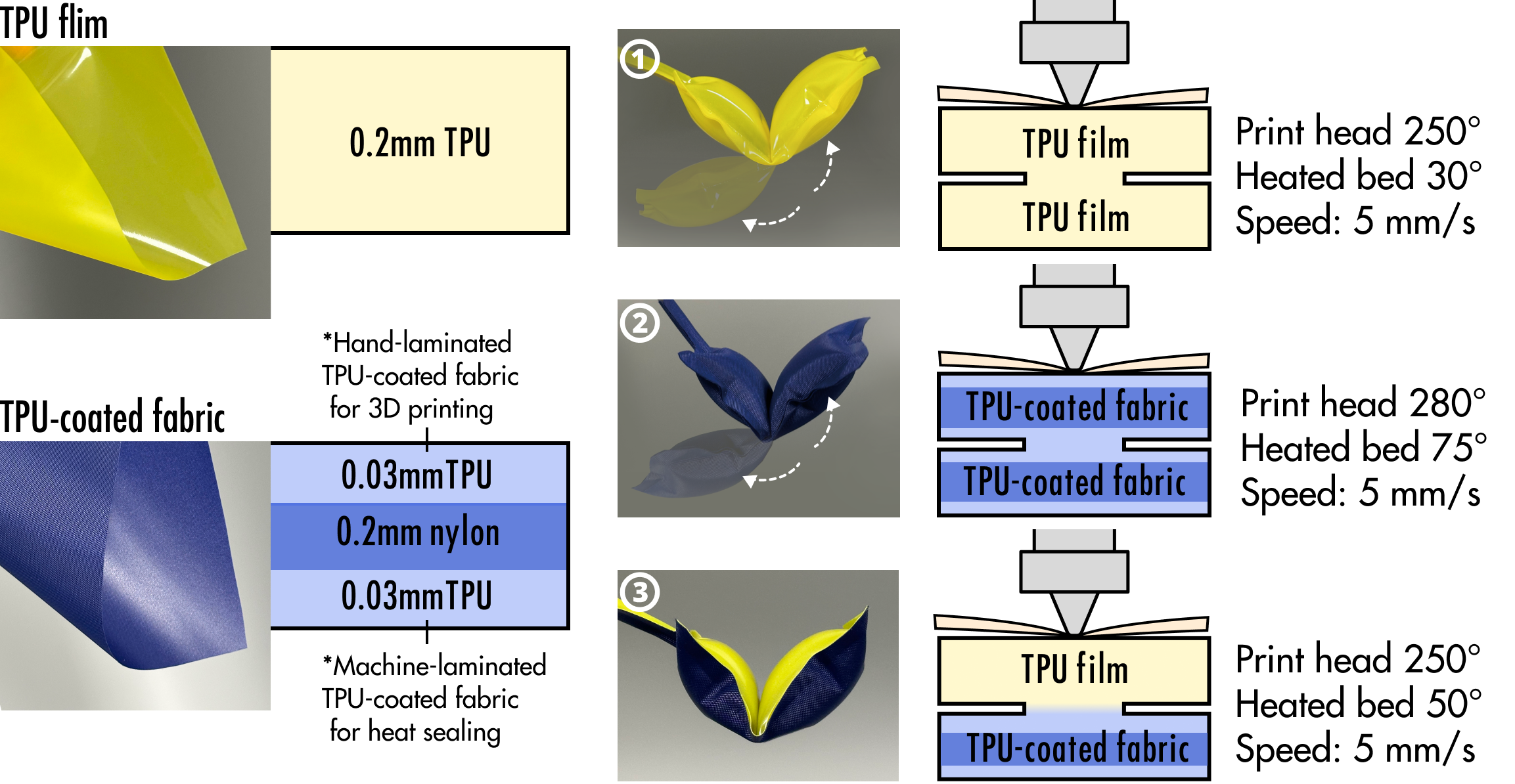}
    \caption{  \textcolor{black}{Materials used for heat-sealing and corresponding settings}.}
    \Description{A comparison of materials and settings for heat-sealing. It shows 0.2mm TPU film and TPU-coated fabric (0.03mm TPU over 0.2mm nylon). Three configurations are depicted: TPU film on TPU film, TPU-coated fabric on TPU-coated fabric, and TPU film on TPU-coated fabric, each with specified print head temperature, heated bed temperature, and speed.}
    \label{fig:fab_film}
\end{figure}
\endgroup

To help users understand the DuoMorph workflow and simplify design and fabrication operations, we developed a GUI design tool based on the Rhino 8 computer-aided design (CAD) tool and the built-in Grasshopper plugin, shown in Fig. \ref{fig:design tool}.a. This tool supports users in converting custom planar sketches into G-code for heat sealing airbags using 3D printers, and merging them with G-code for other components requiring 3D printing to obtain the final complete fabrication G-code.

The specific usage process is as follows:

\begin{enumerate}
    \item \textbf{Define the printing region:} \textcolor{black}{Users first define the printable area in the design tool based on their 3D printer’s build platform dimensions. The tool then displays a corresponding rectangle in Rhino’s workspace as a visual guide (Fig.~\ref{fig:design tool}.b)}.
    
    \item \textbf{Complete the design:} \textcolor{black}{Referring to DuoMorph’s design space, example primitives, and prior work \cite{ou_aeromorph_2016,yang_snapinflatables_2024}, users create heat-sealing patterns and 3D models of printed components in Rhino. If another modeling tool is used, the model can be imported into Rhino for the subsequent steps}.
    
    \item \textbf{Generate heat-sealing G-code:} \textcolor{black}{After design completion, users select the heat-sealing pattern and fabric type in the tool  (Fig.~\ref{fig:design tool}.a,b). The tool automatically converts the sketches into G-code. It samples the paths at 0.5 mm intervals and generates nozzle movements at speeds and heights optimized for the selected fabric. For multi-curve patterns, the nozzle lifts off the fabric between curves to avoid unwanted contact}.
    
    
    \item \textbf{Generate 3D printing G-code using general slicing software:}   \textcolor{black}{Users export the 3D components as STL files via the tool (Fig.  \ref{fig:design tool}.a,c) and slice them using standard third-party slicing software (e.g., Cura or Bambu Studio), which offers reliable settings for FDM printers. Because printing occurs directly on heat-sealed fabric, users must disable features like platform-based supports or brim that could damage the sealed areas. The tool will also generate an alignment marker in the STL to help position parts correctly in the slicing software (Fig. \ref{fig:design tool}.c)}.

    \item \textbf{Merge G-code:} \textcolor{black}{Users return to the design tool and select the sliced 3D printing G-code. The tool automatically combines it with the heat-sealing G-code into a single file. A pause command and MIDI alert sound are inserted after the heat-sealing phase, allowing users to inspect sealing quality and prepare for printing (e.g., by removing heat-insulating PTFE fabric) before continuing}.
    
    \item \textbf{Export and run merged G-code:} \textcolor{black}{Once merged, users can preview the complete G-code in the GUI, save it to a storage device (e.g., an SD card), transfer it to the 3D printer, and initiate the fabrication}.
\end{enumerate}

\subsection{Heat Sealing Materials and Settings}

\subsubsection{Material Selection}

Two commonly used sheet materials for making airbags are adopted in this research: a 0.2 mm TPU film (thermal conductivity$\approx$0.2~W/(m$\cdot$K)) and a 0.2 mm nylon fabric laminated with a 0.03 mm TPU layer (TPU-coated fabric, thermal conductivity$\approx$0.25~W/(m$\cdot$K)) (Fig.~\ref{fig:fab_film}). The TPU film is softer and can therefore be more easily shaped by 4D-printed structures, whereas the TPU-coated fabric is stiffer and can better support printed structures when the airbag is not inflated. Note that most off-the-shelf coated fabrics are single-side laminated, and printed structures adhere poorly to bare nylon. For this reason, it is recommended to laminate the fabric surface intended for printing with an additional TPU film.

When the airbag is sealed using a single type of sheet material, it often, but not always, folds or bends toward the side in contact with the nozzle \cite{ou_aeromorph_2016}. In contrast, when the two materials are combined, the airbag consistently deforms toward the TPU film side, owing to its softer and more stretchable nature.

\subsubsection{Heat-sealing Process (Fig.~\ref{fig:fab_film_setting})}

\begingroup
\setcounter{figure}{10}
\begin{figure}[t]
    \centering
    \includegraphics[width=1\linewidth]{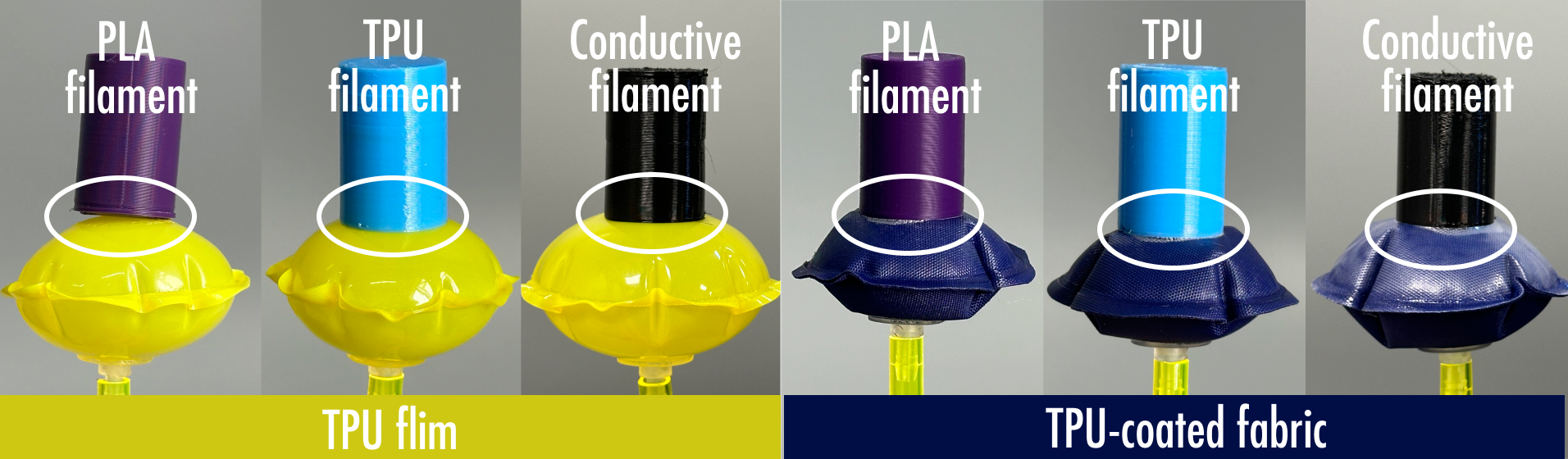}
    \caption{ \textcolor{black}{Qualitative adhesion performance.}}
    \Description{A side-by-side comparison of adhesion performance on two materials: TPU film (yellow) and TPU-coated fabric (blue). For each material, three filaments—PLA (purple), TPU (blue), and conductive (black)—are printed on top. White circles highlight the interface, showing how well each filament adheres to the underlying substrate.}
    \label{fig:fab_filament}
\end{figure}
\endgroup

\begin{figure*}[t]
    \centering
    \includegraphics[width=1\linewidth]{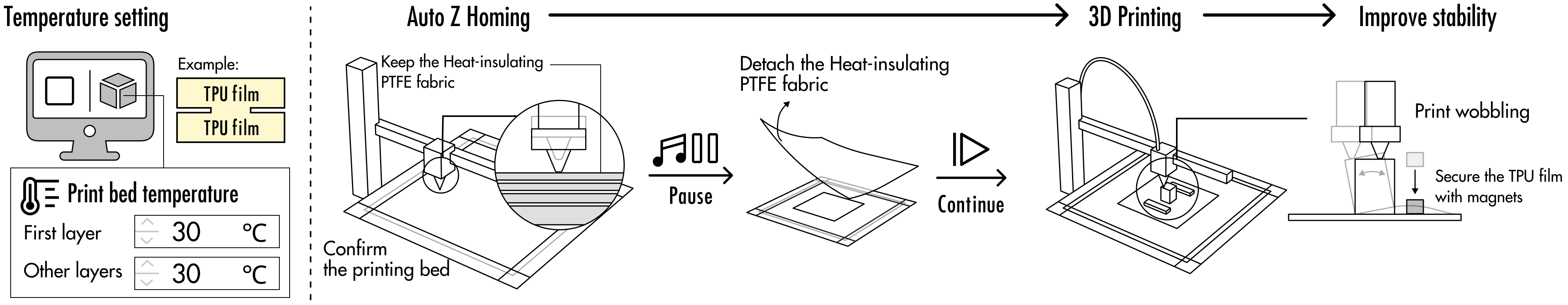}
    \caption{The detailed workflow for FDM printing.}
    \Description{A workflow diagram for FDM printing. It starts with setting the print bed temperature (30°C for all layers). Next, it shows Z-offset adjustment while keeping a heat-insulating PTFE fabric in place. After pausing to detach the fabric, printing continues. The final step addresses print wobbling by securing the TPU film with magnets to improve stability.}
    \label{fig:fab_fdm}
\end{figure*}


\textcolor{black}{Based on the supplier’s recommendations, the nozzle temperature was set to 250 ℃ or 280 ℃ and the heated-bed temperature to 50 ℃ or 70 ℃, both determined by the top-layer fabric. The print speed was fixed at 5 mm/s to ensure reliable heat sealing during printing (Fig.~\ref{fig:fab_film})}.
 In addition, a protective PTFE fabric \textcolor{black}{(0.1mm thick, thermal conductivity: 0.23~W/(m$\cdot$K))} is a applied when sealing TPU film to prevent scratching of the thin film. The PTFE fabric is as required for TPU-coated fabric to protect the coating. \textcolor{black}{Finally, the force exerted during Z-axis homing at printer initialization is found to be sufficient for heat sealing. and therefore no additional press distance is specified in the G-code to increase force. Moreover, variations in film or fabric thickness can be accommodated by the homing process, eliminating any need for manual adjustment.}

\subsection{FDM Printing Filaments and Settings}

\subsubsection{Filament Selection}

Three types of filament are used in this research: \textcolor{black}{PLA (Bambu), TPU (Bambu), and TPU-based conductive filament (Qie feng)}. Among them, PLA and TPU are the most frequently applied. Structures printed with TPU generally exhibit stronger bonding with TPU films, owing to their chemical and mechanical compatibility (Fig.~\ref{fig:fab_filament}, left). Both TPU and PLA can adhere reasonably well to TPU-coated fabric, likely because the fabric layer reduces the elasticity compared to pure TPU film (Fig.~\ref{fig:fab_filament},right). Quantitative results of the bonding strength will be presented in Section~\ref{sec:adhension_test}. The conductive filament is primarily employed to introduce sensing capability.

\begin{figure}[b]
    \centering
    \includegraphics[width=1\linewidth]{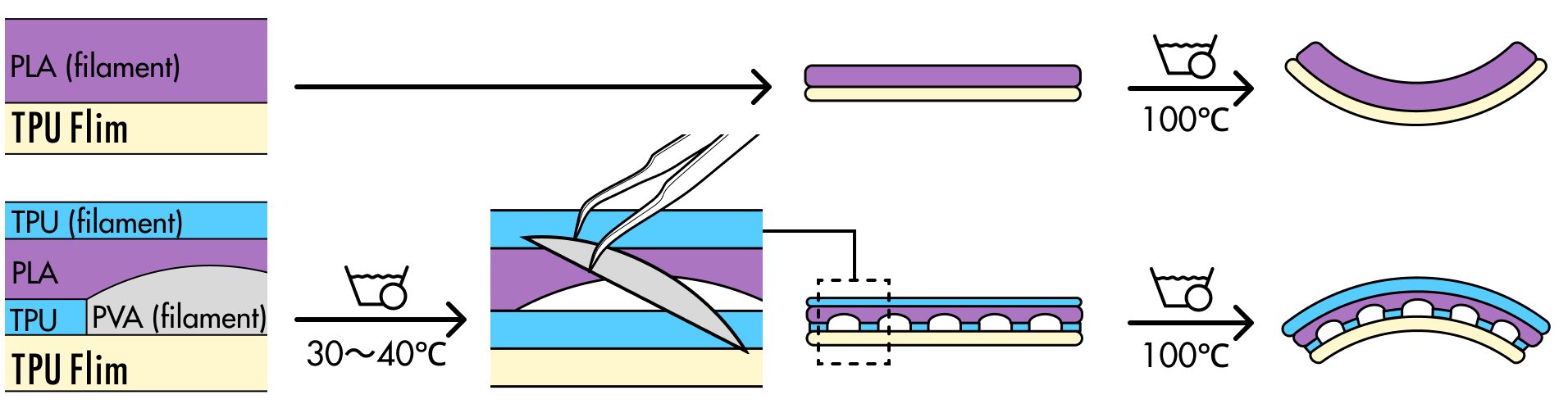}
    \caption{Additional treatment steps for 4D printing.}
    \Description{A diagram illustrating additional treatment steps for 4D printing. It shows two layered structures: one with PLA over TPU film, and another with TPU/PLA/PVA over TPU film. The second structure is treated at 30–40°C to remove the PVA layer, creating a patterned interface. Both structures are then exposed to 100°C water, causing them to bend into curved shapes.}
    \label{fig:fab_4d}
\end{figure}

\begin{figure}[b]
    \centering
    \includegraphics[width=1\linewidth]{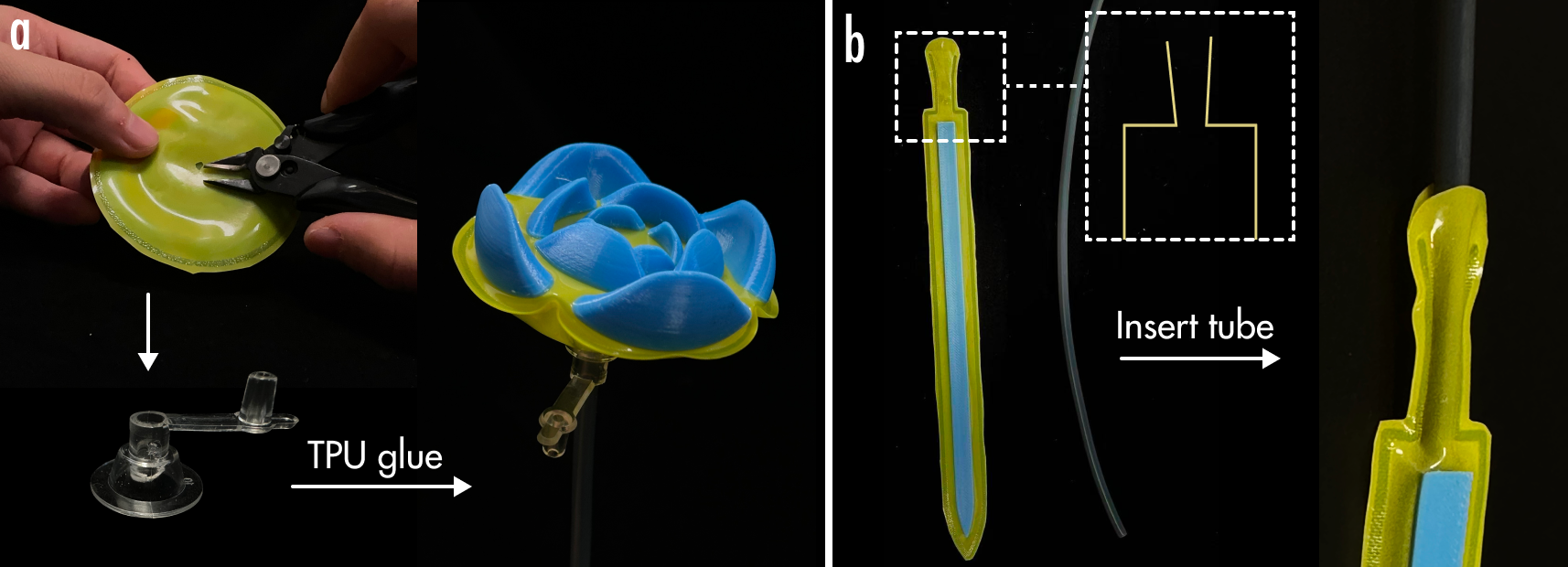}
    \caption{  \textcolor{black}{Two tubing assembly approaches.}}
    \Description{Two approaches for assembling tubing onto printed parts. Part (a) shows a hand using pliers to attach a tube to a yellow component, followed by an illustration of using TPU glue to secure a clear connector. Part (b) displays a yellow sword-shaped part with a pre-printed channel for tubing, which is then inserted into the channel.}
    \label{air channel}
\end{figure}

\subsubsection{FDM process}
In this step (Fig.~\ref{fig:fab_fdm}), a key difference from conventional 3D printing is that a heated print bed may soften the sheet materials—especially the TPU film—resulting in an unstable substrate and poor print quality. Therefore, it is recommended to keep the bed at room temperature (i.e., turn off bed heating), which our design tool automatically sets in the exported G-code. \textcolor{black}{Except for setting the print bed temperature to 30 °C to prevent thermal deformation of the pneumatic actuator, all other parameters can follow the default recommendations of Bambu Studio for standard 3D printing. For 4D printing, configurations such as toolpath and print speed can be set according to previous work. \cite{an_thermorph_2018,wang_-line_2019}}.

Another point of attention arises in 4D printing is, when printing structures for concave bending, it is advisable to add dissolvable supports for the arch, particularly when the span is large (Fig.~\ref{fig:fab_4d}). After printing, the support should first be dissolved in warm water (30–40 °C) before placing the structure in hot water to trigger deformation. Quantitative experiments to determine the optimal designs of the arch structures will be presented in Section~\ref{sec:4d_test}. In addition, when the base of the arch structure in contact with the sheet material is narrow, it is recommended to add a thin intermediate TPU layer to enhance the interfacial adhesion between the PLA arch and the sheet material.

\subsection{Post-processing}
\textcolor{black}{After heat-sealing and printing, the actuator can be trimmed and the tubing manually assembled. The tubing can be assembled either vertically or horizontally. For vertical assembly (Fig.~\ref{air channel}.a), a hole is cut in the actuator, and a valve is glued in place. The tube is then connected to the valve. Alternatively, if a channel was reserved during heat-sealing, the tube can simply be inserted horizontally and secured with adhesive (Fig.~\ref{air channel}.b)}.

\section{Applications}
\begin{figure}[b]
    \centering
    \includegraphics[width=1\linewidth]{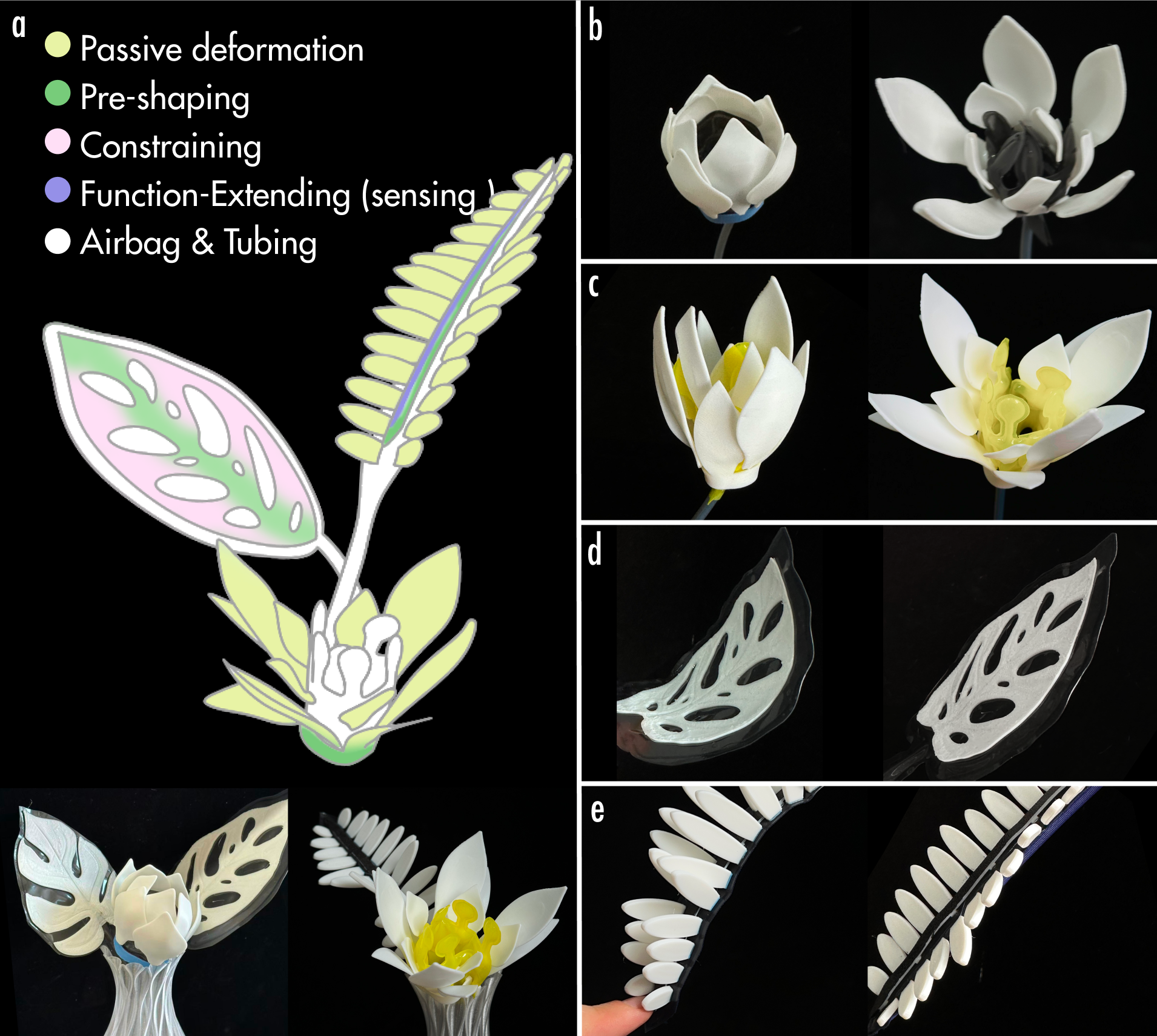}
    \caption{Kinetic sculpture integrating multiple types of Duomorph structures.}
    \Description{A kinetic sculpture combining multiple Duomorph structures. Panel (a) provides a legend for components: passive deformation (yellow), active deformation (green), constraining (pink), function-extending sensing (blue), and airbag & tubing (white). Panels (b-e) show different views of the white flower-like sculpture, demonstrating its unfolding motion and structural details, including leaves and petals with integrated sensors.}
    \label{fig:app-sculpture}
\end{figure}
To demonstrate the potential of Duomorph, we designed a set of application artifacts, including a kinetic sculpture, a biomimetic gripper, a desktop toy, and a massage neck pillow, each showcasing how different DuoMorph structures can be integrated to achieve expressive, functional, and interactive shape-changing interfaces.

\subsection{Kinetic Sculpture}

This example leverages all four types of printed structures to create plant-like kinetic sculptures that interact with visitors (Fig.~\ref{fig:app-sculpture}.a). The flowers are equipped with 4D printed strips at their bases, allowing them to curl into cylinders when heated. The petals, printed directly on airbags, bloom when inflated (Fig.~\ref{fig:app-sculpture}.b,c). A large leaf incorporates a 4D printed surface that bends the leaf after heated but flattens once the air chamber inflates (Fig.~\ref{fig:app-sculpture}.d). The mimosa-inspired leaf integrates a 4D printed stem that curves a little bit to shape the leaf after heated and contains a sensing layer to detect human touch (Fig.~\ref{fig:app-sculpture}.e). When touched, all airbags deflate, causing the mimosa leaves to fold, flowers to close, and large leaves to curve.

\subsection{Biomimetic Gripper}

\begin{figure}[b]
    \centering
    \includegraphics[width=1\linewidth]{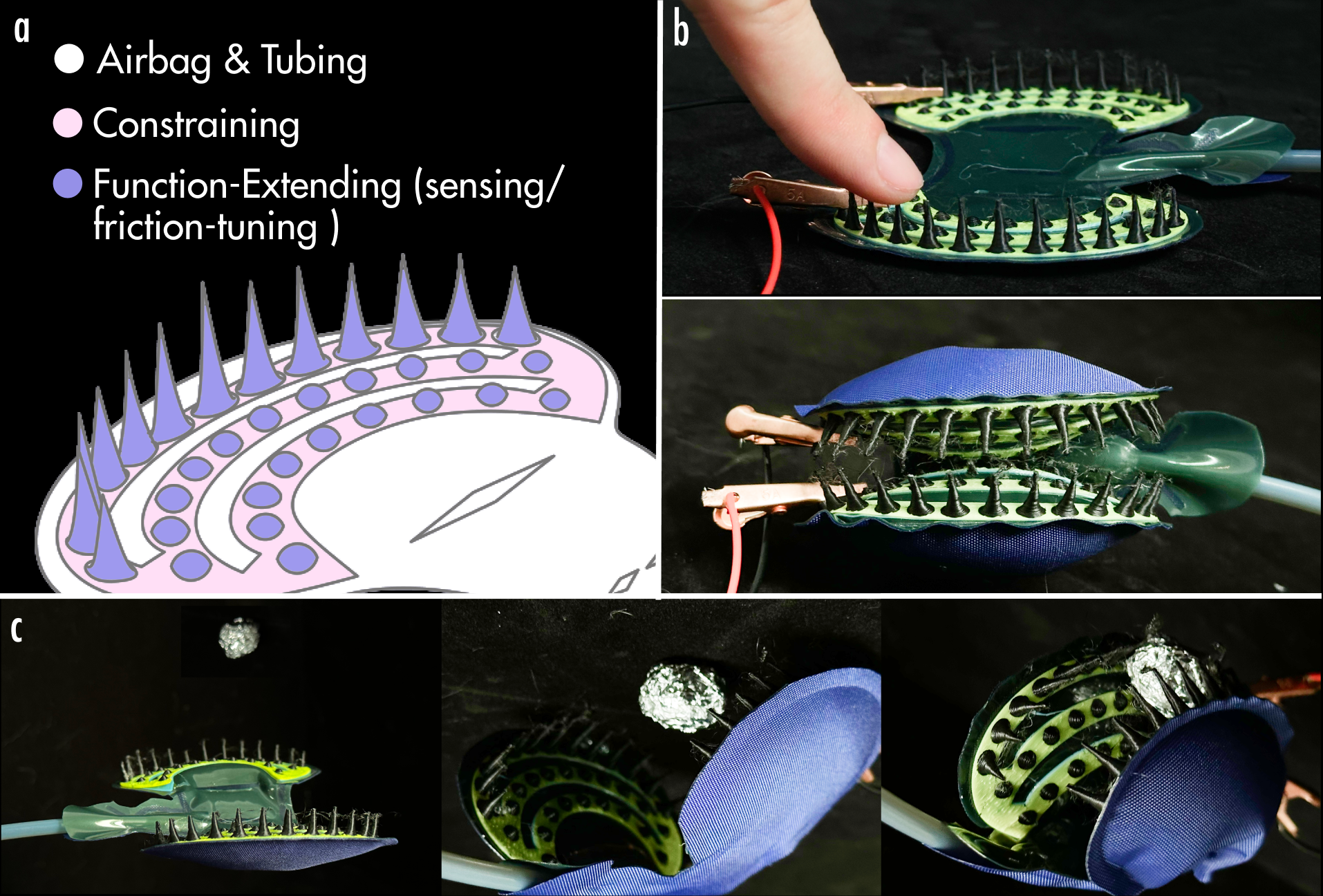}
    \caption{Biomimetic gripper inspired by \textit{Dionaea muscipula} (Venus flytrap).}
    \Description{A biomimetic gripper inspired by the Venus flytrap. Panel (a) is a diagram showing its layered structure: airbag & tubing (white), constraining layer (pink), and function-extending elements for sensing/friction-tuning (blue). Panels (b) and (c) show the physical device—a green, spiky, clamshell-like gripper—closing on an object and gripping it securely.}
    \label{fig:app-gripper}
\end{figure}

This gripper is inspired by the Venus flytrap and is enabled by printed constraint, sensing, and friction-tuning structures (Fig.~\ref{fig:app-gripper}.a). A surface coating constrains air deformation when inflated, while dot-shaped protrusions increase friction during contact. In addition, spike-like structures printed with conductive PLA filament serve as sensors. As shown in Fig.~\ref{fig:app-gripper}.b, the gripper can respond to human touch. Moreover, when in contact with conductive objects (e.g., an aluminum foil ball), the sensors trigger the gripper to close around the object (Fig.~\ref{fig:app-gripper}.c).

\subsection{Customized Massage Neck Pillow}
\begin{figure}[b]
    \centering
    \includegraphics[width=1\linewidth]{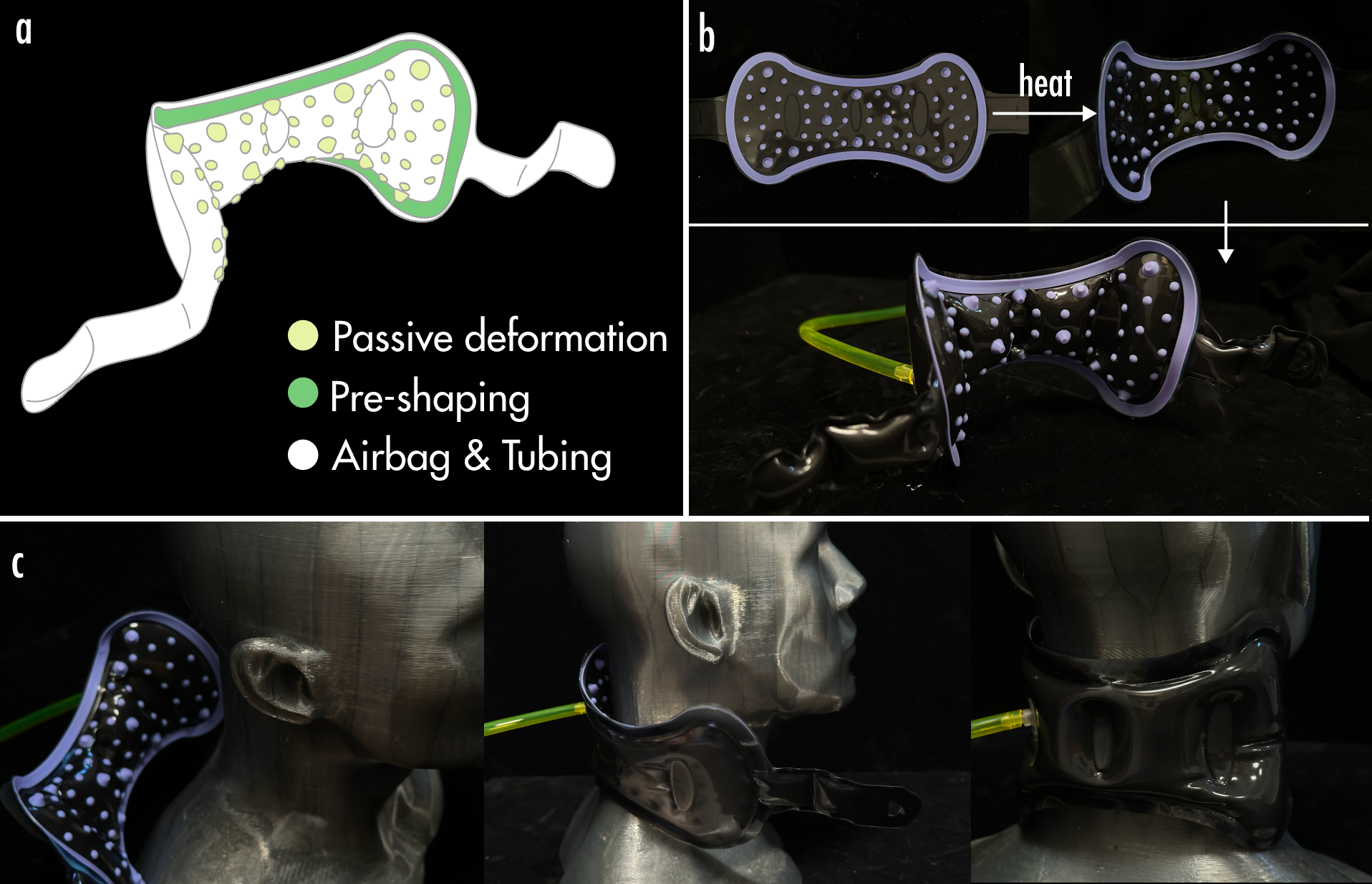}
    \caption{A customized neck pillow for Massage.}
    \Description{A customized neck pillow for massage. Panel (a) is a diagram showing its structure: passive deformation (yellow dots), active deformation (green outline), and airbag & tubing (white). Panel (b) shows the pillow inflating when heated. Panel (c) displays the pillow fitted onto a mannequin head and neck, demonstrating its conforming shape for ergonomic support.}
    \label{fig:app-pillow}
\end{figure}

\textcolor{black}{This neck pillow (Fig.~\ref{fig:app-pillow}.a) is pre-shaped by a 4D-printed frame, which can be customized to the user’s neck dimensions for optimal fit and conformity (Fig.~\ref{fig:app-pillow}.b,c). By embedding the geometric curvature directly into the 4D-printed structure, the pillow naturally aligns with the user’s cervical contour without requiring continuous actuation. In addition, the 4D-printed frame offloads the shaping and support functions from the pneumatic actuator, allowing the actuator to focus exclusively on driving the inflation of the printed massage dots. This division of roles not only simplifies the actuator design but may also enhance the overall massage performance by enabling more targeted and consistent actuation.}

\subsection{Desktop Toy}
\begin{figure}[b]
    \centering
    \includegraphics[width=1\linewidth]{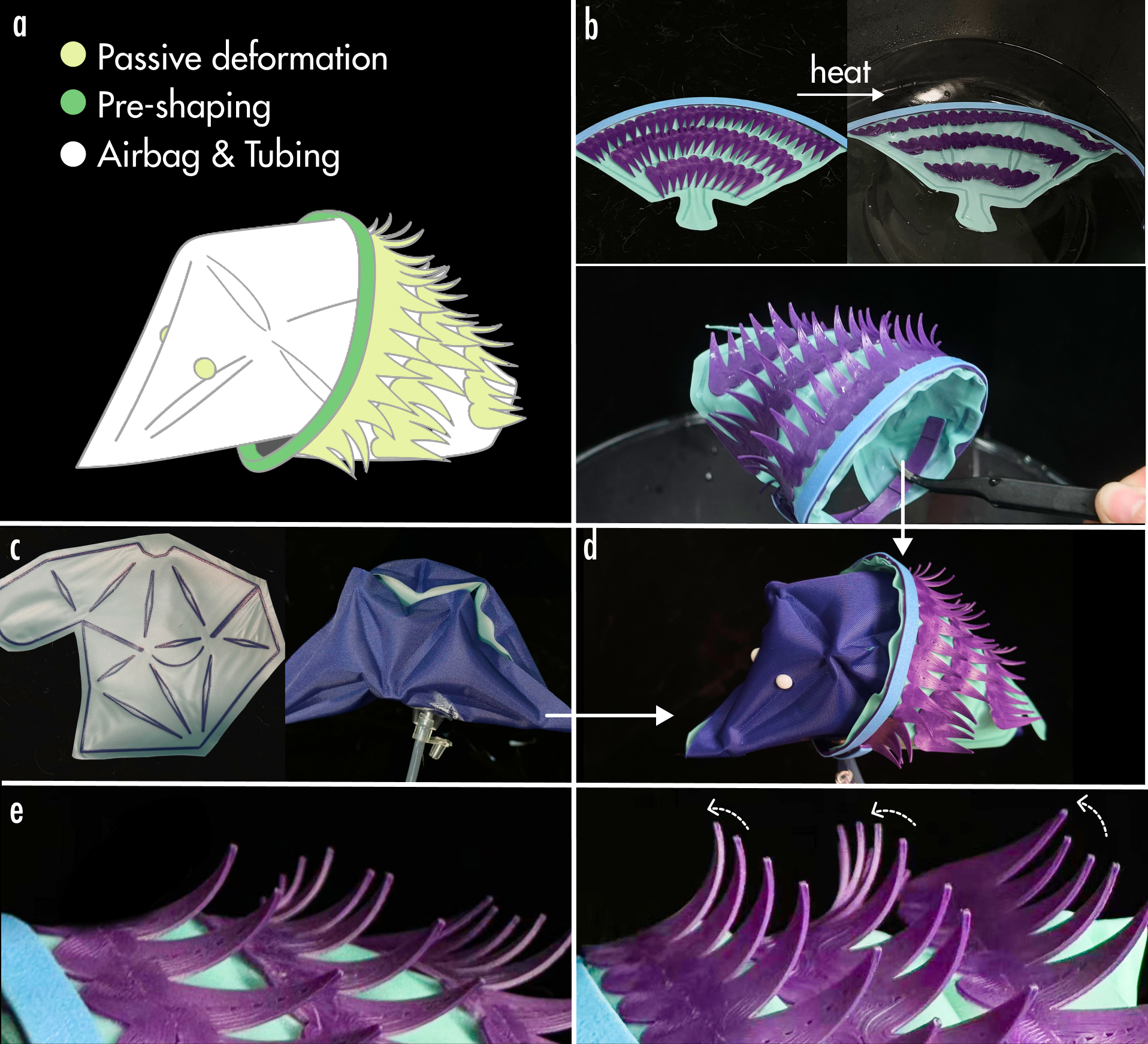}
    \caption{Hedgehog-inspired desktop toy demonstrating dynamic actuation.}
    \Description{A hedgehog-inspired desktop toy demonstrating dynamic actuation. Panel (a) is a diagram showing passive deformation (yellow), active deformation (green), and airbag & tubing (white) components. Panels (b-e) show the toy’s transformation: heating causes its purple spines to curl, then a hand inserts a tube to inflate the body, causing the spines to extend outward dynamically.}
    \label{fig:app-toy}
\end{figure}

By integrating printed pre-shaping and passive deformation with pneumatic actuation, we created a hedgehog-inspired desktop toy (Fig.~\ref{fig:app-toy}.a). The toy consists of two main components: an inner inflatable body and an outer 4D-printed shell that forms both the skin and the spines. When heated, the outer shell undergoes a programmed shape transformation, rolling into a cone-like geometry while the originally flat spines gently lift upward (Fig.~\ref{fig:app-toy}.b). Once mounted on the inflated inner body, the shell can further expand, causing the spines to stand more prominently and enhancing the overall expressive effect. Through repeated inflation and deflation cycles, the toy produces rhythmic swaying and pulsating motions, giving rise to a playful, engaging interaction reminiscent of a small creature’s breathing or movement (Fig.~\ref{fig:app-toy}.c). 

\begin{figure*}[t]
    \centering
    \includegraphics[width=1\linewidth]{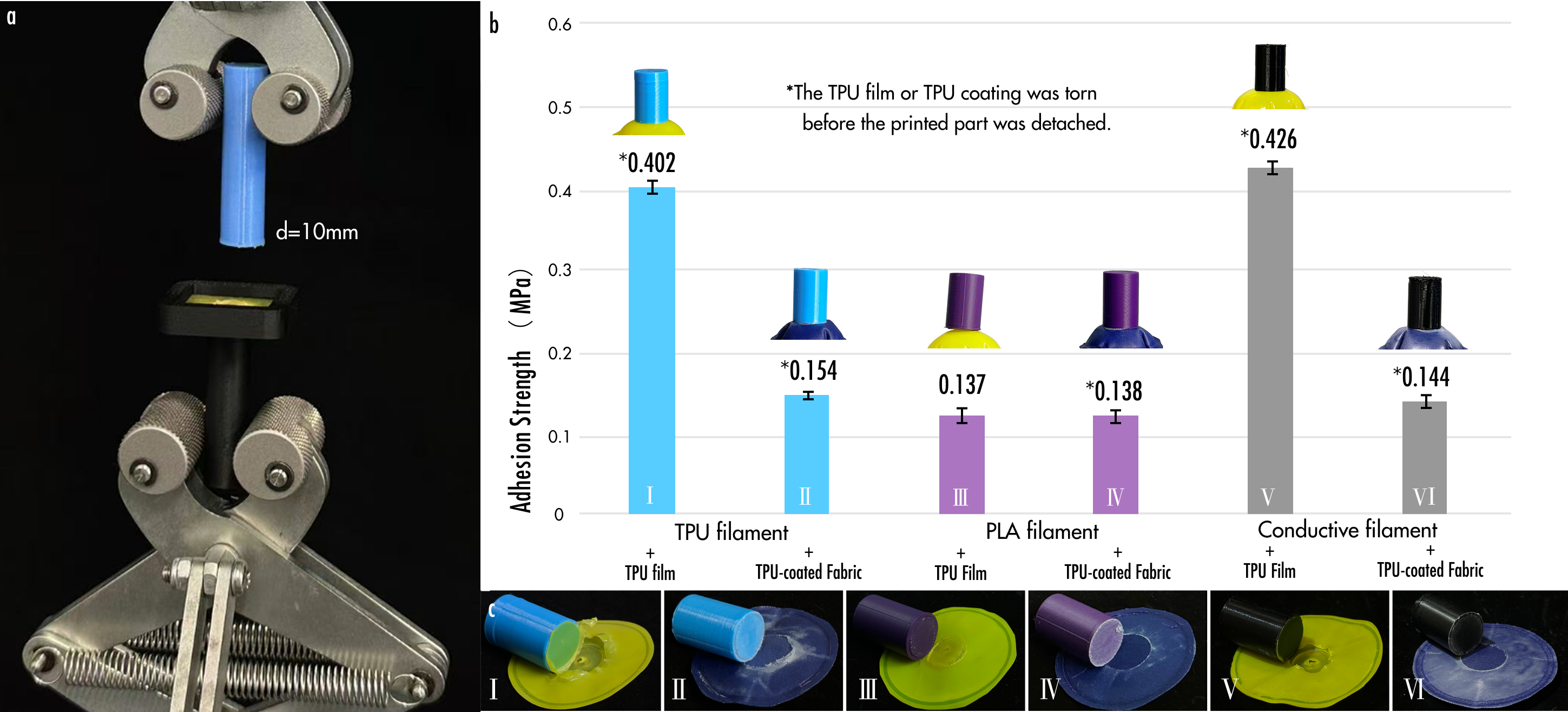}
    \caption{\textcolor{black}{Tensile strength measurements for TPU film, TPU-coated fabric, and PLA/TPU combinations. (a) Test setup, (b) Test restults, (c) post-separation contact surfaces across experimental groups}.}
    \Description{Adhesion strength measurements. Panel (a) shows the experimental setup using a tensile tester to pull a printed cylinder (d=10mm) from a film substrate. Panel (b) is a bar chart comparing adhesion strength (MPa) across six material combinations. Panel (c) displays photos of the post-separation surfaces, with asterisks indicating cases where the TPU film or coating tore before detachment.}
    \label{fig:adhesion}
\end{figure*}

\begin{figure}[t]
    \centering
    \includegraphics[width=1\linewidth]{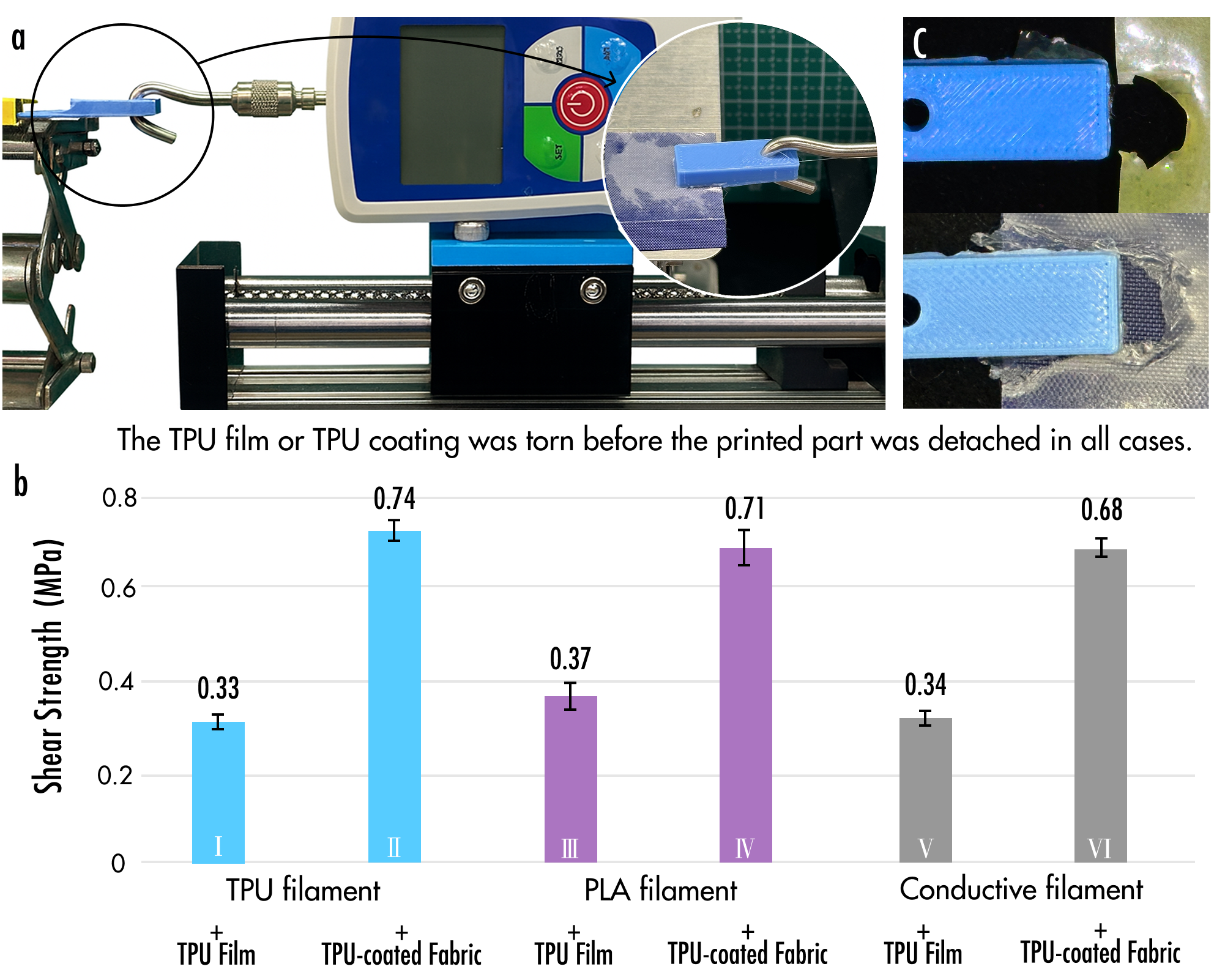}
    \caption{ \textcolor{black}{Shear strength measurements for TPU film, TPU-coated fabric, and PLA/TPU combinations. (a) Test setup, (b) Test restuls. (c)  The TPU film and TPU coating were torn before the printed part was detached in all these six cases.}}
    \Description{Shear force tests. Panel (a) shows the test setup: a digital force gauge pulling a printed sample bonded to a film substrate on a linear rail. Panel (b) is a bar chart displaying shear adhesion strength (MPa) for six material combinations. An asterisk notes that in all six cases, the TPU film or coating tore before the printed part detached.}
    \label{Shear Force}
\end{figure}

\section{Evaluation}

\subsection{Adhesion Strength}\label{sec:adhension_test}
The adhesion strength between printed structures and sheet materials is one of the most critical parameters in Duomorph.
\textcolor{black}{To evaluate the tensile strength, we printed cylindrical samples (diameter: 10\,mm, height: 40\,mm) on different sheet materials vertically. Then each combination was tested ten times using a universal testing machine.} \textcolor{black}{As shown in Fig.~\ref{fig:adhesion}, TPU film combined with regular TPU filament and TPU-based conductive filament (groups i and v) exhibited the highest tensile strength. In fact, the actual bonding strength is even greater, as the TPU film itself ruptured prior to separation. In contrast, TPU film or coating combined with PLA filament (groups iii and iv) demonstrated the weakest tensile strength. And TPU‑coated fabric combined with TPU filaments (groups ii and vi) also exhibited relatively low tensile strength due to the weaker adhesion between the hand‑laminated TPU coating and the nylon substrate. These results indicate that materials of the same type exhibit significantly stronger interfacial adhesion, leading to improved mechanical performance}.

\begin{figure*}[t]
    \centering
    \includegraphics[width=1\linewidth]{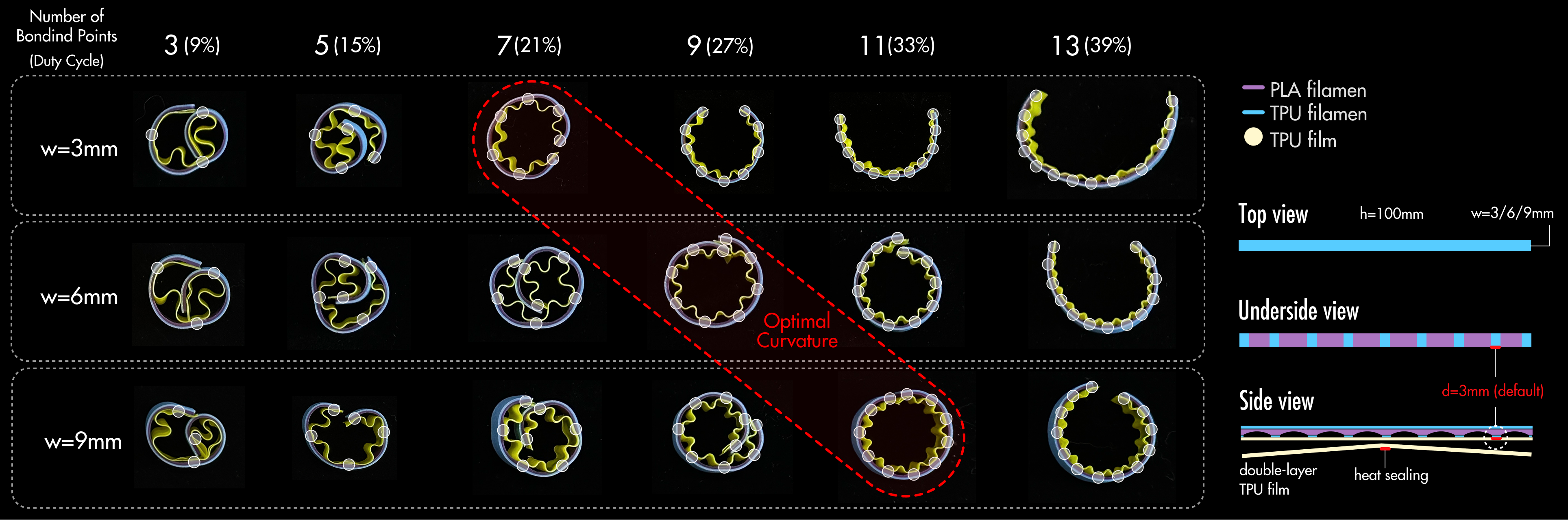}
    \caption{Bending curvature results under different strip widths and bonding ratios. Red dashed line marks optimal circular bending.}
    \Description{A grid showing bending curvature results for strips with different widths (3mm, 6mm, 9mm) and bonding point counts (3 to 13). Each cell displays a printed strip’s final curved shape. A red dashed line highlights the “Optimal Curvature,” indicating the combination of width and bonding ratio that produces the most circular bend.}
    \label{fig:4d-bending}
\end{figure*}

\begin{figure}[t]
    \centering
    \includegraphics[width=1\linewidth]{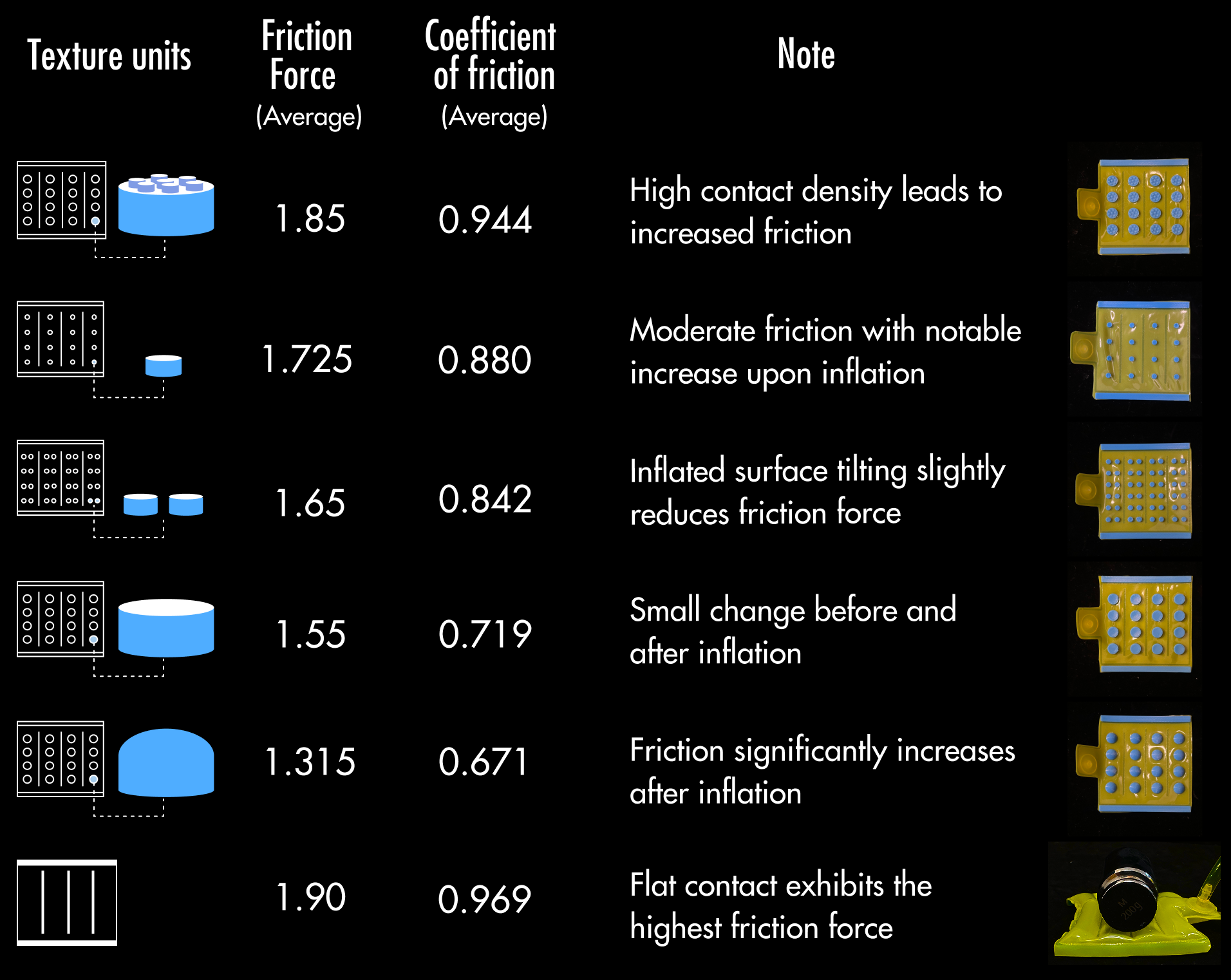}
    \caption{Comparison of friction coefficients for different printed surface textures and structures.}
    \Description{A table comparing friction coefficients for six different printed surface textures. Each row shows a texture unit diagram, its average friction force and coefficient, a note on its behavior (e.g., “High contact density leads to increased friction”), and a photo of the physical sample. Results show how inflation and surface geometry affect friction.}
    \label{fig:friction}
\end{figure}

\textcolor{black}{Shear force tests were conducted using a shear force tester, as shown in Fig.~\ref{Shear Force}.a. Thin brick-shaped samples (20 mm × 10 mm × 4 mm) were printed across the edge of different sheet materials, and each combination was tested ten times. Fig.~\ref{Shear Force}.b shows the results. The actual shear strength will be even greater, as the TPU films or coatings ruptured before the printed parts detached in all six combinations (Fig.~\ref{Shear Force}.c). This explains the similar results observed within groups i, iii, v and groups ii, iv, vi, which primarily reflect the force required to tear the TPU film or peel the TPU coating from the nylon substrate.}

For TPU-coated fabric, which is originally single-sided, the TPU layer between the fabric and the filament was produced through manual lamination. Although the manually laminated TPU layer was thin (0.03\,mm), the TPU filament penetrated into the woven nylon gaps during thermal pressindue to minor plastic deformationg, forming both micro-mechanical interlocks and molecular-level wetting. As a result, the adhesion strength was higher for both TPU and PLA printed structures compared to plain TPU film.

\subsection{Concave Bending in 4D Printing}\label{sec:4d_test}
To investigate factors affecting concave bending curvature, we prepared three sets of samples (widths: 3, 6, and 9\,mm; length: 100\,mm) printed on double-layer TPU films. For each width, six bonding ratio patterns (bonding point width: 3\,mm) were tested to analyze the relationship between bonded area and curvature. Other variables such as water temperature (100$^{\circ}$C), material, and fabrication process were kept constant.

As shown in Fig.~\ref{fig:4d-bending}, the results indicate:
\begin{enumerate}
    \item Lower bonding ratios (larger unbonded regions) produced stronger bending, until limited by the curling stiffness of the TPU film.
    \item At the same bonding ratio, wider strips resulted in greater curvature.
    \item The optimal bending condition (approximating a closed circle, marked with a red dashed line) followed a diagonal trend: as width increased, the required number of bonding points also increased to maintain similar bending efficiency.
\end{enumerate}

In summary, for large-area pneumatic actuators, wider concave-bending strips are recommended to enhance curvature. For small-area actuators, the design can be tuned more flexibly:
\begin{itemize}
    \item To achieve strong curvature, use low bonding ratios or wider strips.  
    \item To achieve mild curvature, use high bonding ratios or narrower strips.  
    \item To form a near-perfect circular arc, select the appropriate combination of width and bonding ratio along the optimal curvature trend.  
\end{itemize}

\subsection{Friction Tunability}\label{sec:friction_test}

We also evaluated how different surface patterns influence frictional performance (results shown in Fig.~\ref{fig:friction}). All samples are fully inflated and is tested with a 200g weight loaded. In general, adding printed textures reduced the measured coefficient of friction, likely due to decreased effective contact area. Alternativly, other type geometries—such as flat surface patches or suction-cup-like features—might significantly increase friction and adhesion. These results highlight the design flexibility of Duomorph in tailoring surface interaction properties for specific applications.

\section{Limitations, Discussion, and Future Work}

\textbf{Single-Sided Printing.}  
In the current workflow, structures can only be printed on one side of the airbag, which introduces constraints that must be carefully considered during the design phase. For example, achieving convex 4D bending requires specifically designed bases to redirect the deformation to the printable side. Another strategy is to leverage modular connectors so that multiple prototypes can be joined back-to-back, enabling both sides of the combined structure to host functional features.  

\textbf{Printing Height Limits on Unstable Substrates.}  
Replacing the conventional rigid build plate with a soft film introduces challenges in height control. During printing, deformation of the flexible substrate can cause vertical displacement, leading to misaligned layers or nozzle collisions. This issue is particularly evident when using TPU film as the actuator material, as heating or pressure changes can lift the surface. (In contrast, TPU-coated fabric, due to its stiffness, shows minimal height variation.) To mitigate this, we fixed the actuator to the platform with double-sided tape to suppress displacement and reduced print speed to lower mechanical impact. With these measures, we successfully printed TPU filaments up to 7\,cm tall on TPU film substrates. Looking ahead, vacuum-based clamping systems specialized for thin-film materials (e.g., \cite{lu_millimorph_2019}) could be adapted to clamp the films to further stabilize the process.  

\begin{figure}[t]
    \centering
    \includegraphics[width=1\linewidth]{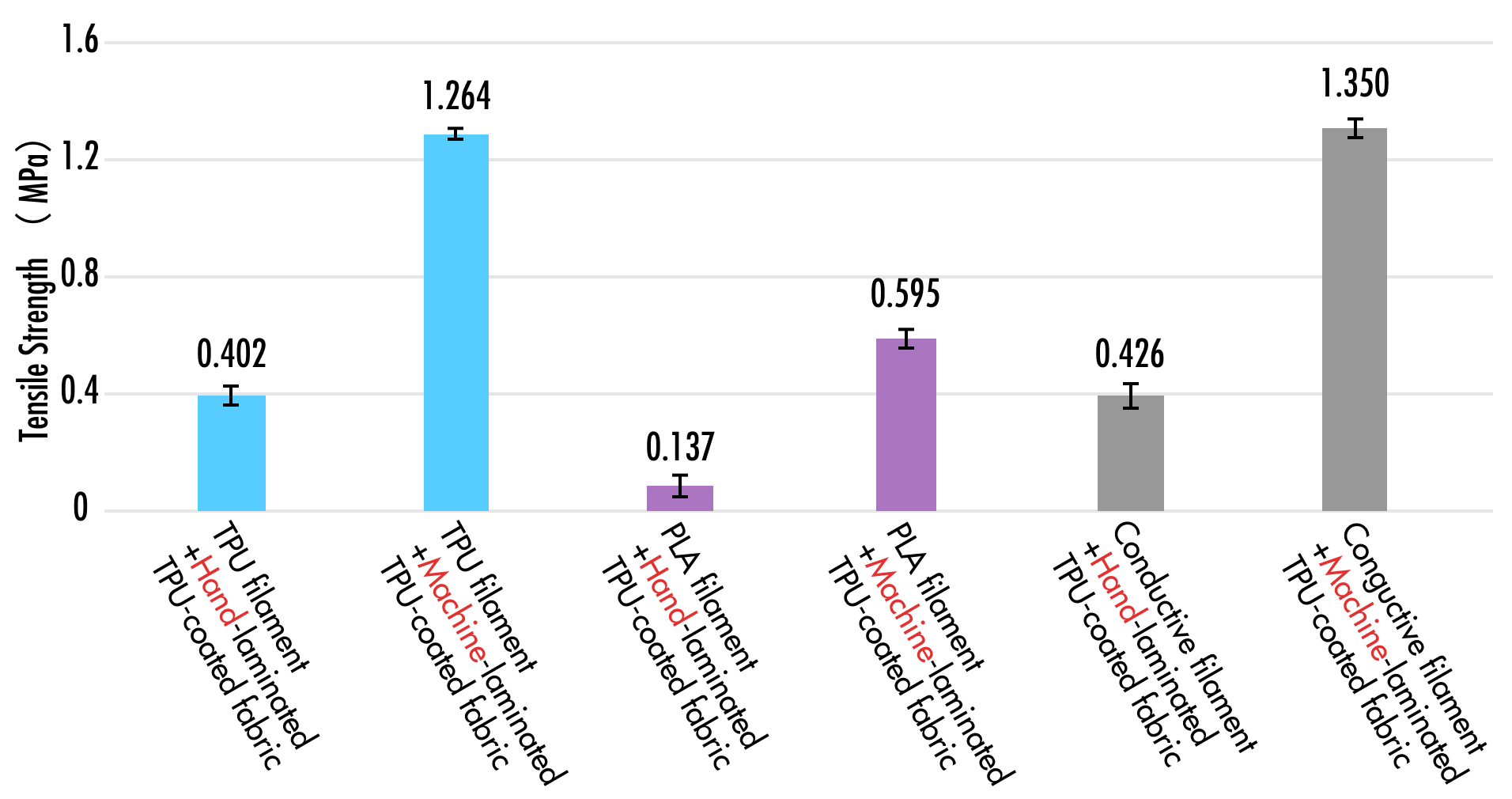}
    \caption{  \textcolor{black}{ Adhesion strength comparison between manually coated and industrially coated double-sided TPU fabrics}.}
    \Description{ A bar chart comparing adhesion strength (MPa) between manually coated and industrially coated double-sided TPU fabrics. For each filament type (TPU, PLA, Conductive), the machine-laminated fabric shows significantly higher adhesion strength than the hand-laminated version, demonstrating the benefit of industrial processing.}
    \label{fig:adhesion-test2}
\end{figure}

\textbf{Improved Double-Sided TPU-Coated Fabric.}  
Most commercially available TPU-coated fabrics are single-side coated, requiring manual lamination of an extra TPU layer before FDM printing (since nylon surfaces are non-adhesive to printed polymers). Heat pressing or ironing TPU film onto textiles is possible, but manual coating is highly quality-sensitive: trapped air pockets or uneven pressure often produce weak bonds prone to delamination during printing. To overcome this, we sourced machine-laminated double-sided TPU-coated fabrics ($\sim$0.5\,mm thick, about twice the thickness of the single-sided variant we currently use). In a comparative tensile strength test, the bonding strength with TPU filament increased by approximately 3.2 times, while that with PLA filament increased by about 4.3 times ( Fig.~\ref{fig:adhesion-test2}).

\textbf{Structural Reinforcement via Weaving.}  
During our research, we found that TPU filament exhibits superior interfacial bonding with TPU film, and is therefore often used as the first printed layer to improve adhesion of rigid filaments such as PLA. However, simple layer stacking remains vulnerable to delamination under concentrated stress. To address this, we propose weaving-inspired reinforcement: interlacing print paths or embedding interlocking regions to create mechanical interlocks across layers and interfaces. This strategy has the potential to enhance tensile and shear resistance, leading to stronger and more durable hybrid structures.  

\textbf{Multi-material support, generalization and durability.}  
\textcolor{black}{Compared to SLA printing, FDM printing presents a lower barrier for multi-material fabrication and, by nature, SLA does not support heat sealing. In this work, we explore two types of fabrics combined with a range of printing filaments, enabling varied bonding outcomes. Empirically, materials with similar properties tend to exhibit stronger adhesion. Prior studies \cite{hackaday_fusing_2020, choi_therms-up_2021} have begun to explore FDM-based thermal bonding; however, they still present minor issues, such as uneven finished surface. DuoMorph advances heat-sealing quality by incorporating a protective PTFE fabric, achieving stable and smooth thermal bonding with FDM printers. This smooth and reliable bonding surface further supports high-quality subsequent FDM printing directly onto the fabric. These efforts make the technology theoretically more generalizable across different FDM printers.
In addition, durability tests demonstrate excellent performance of DuoMorph prototypes. The airtightness of the seams is outstanding: TPU-film-based airbags ruptured at the film before seam failure, while TPU-coated fabric airbags withstood the maximum air pressure of the testing machine without leakage. In another test, samples underwent 1,000 inflation-deflation cycles and remained fully functional, with no detachment of printed structures and no leakage. Detailed results are provided in the Appendix.}


\section{Conclusion}\label{sec:discussion}

In this work, we introduced \textit{DuoMorph}, a design and fabrication approach that integrates FDM printing with pneumatic actuation to realize novel shape-changing interfaces. By co-designing printed structures and heat-sealed pneumatic elements, DuoMorph enables actuation and constraint mechanisms that are difficult for either modality to achieve alone. Importantly, the entire hybrid structure can be produced in a single, seamless workflow using a standard FDM printer, encompassing both heat-sealing and 3D/4D printing. We formalized the design space through four primitive categories that characterize the fundamental modes of interaction between printed and pneumatic components. In addition, we developed a fabrication method and supporting design tool to operationalize this process. Through example applications and performance demonstrations, we showcased the versatility and expressive potential of DuoMorph, \textcolor{black}{ranging from kinetic sculptures and biomimetic grippers to playful interactive artifacts and functional neck pillows }. Together, these contributions establish DuoMorph as a new pathway for creating hybrid material systems that expand the possibilities of shape-changing interfaces, and lay the groundwork for future explorations in responsive, functional, and interactive design.

\begin{acks}
We acknowledge Hengrong Ni’s assistance with rapid early‑stage validation when the core concept was first proposed by Dr. Lu. We thank the MiLab at Tsinghua University for its support, and Prof. Huirong Le of The Future Laboratory for providing equipment for performance evaluation. Special thanks go to Tingyu Hua from Prof. Le’s team for his assistance with the tests. We also thank Jeremy Chen from the Morphing Matter Lab for his early coding support. This work was supported by the PolyU Strategic Hiring Scheme (Project P0059979) and partially by the U.S. National Science Foundation (Award No. 2427455).
\end{acks}

\balance
\bibliographystyle{ACM-Reference-Format}
\bibliography{FDM-references-yty}
\balance

\begin{figure*}[t]
\begin{minipage}[b]{\linewidth}
    \centering
\includegraphics[width=1\textwidth]{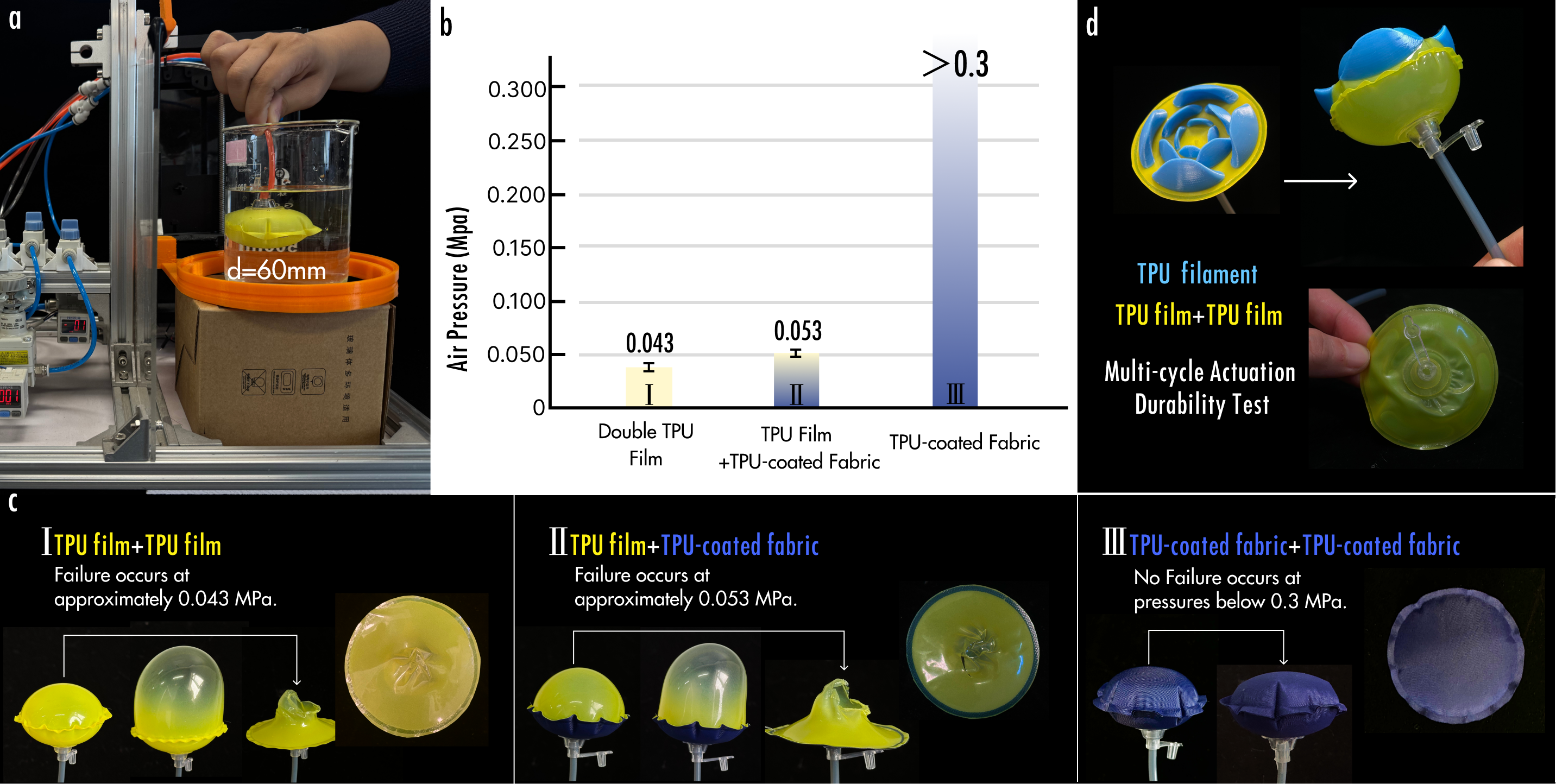}
    \caption{\textcolor{black}{Airtightness and resuability experiments. (a) Test setup, (b) Airtightness test results, (c) Tested airtightness sample details, (d) A tested reusability sample details.}}
    \Description{Airtightness and reusability experiments. Panel (a) shows the test setup: a sample submerged in water while being pressurized. Panel (b) is a bar chart showing failure pressure for three material combinations, with TPU-coated fabric exceeding 0.3 MPa. Panels (c) and (d) show photos of the tested samples before and after failure, illustrating their durability under repeated inflation cycles.}
    \label{appendix}
\end{minipage}
\end{figure*}

\newpage

\appendix
\section{Durability Test}
\subsection{Airtightness}
\textcolor{black}{The airtightness of three types of pneumatic actuators, fabricated by combining two types of film/fabric, was tested using the setup shown in Fig.~\ref{appendix}.a. Air leakage was detected through visual observation of bubbles and by monitoring the air pressure drop. As illustrated in Fig.~\ref{appendix}.b and c, Groups 1 and 2 failed due to rupture of the TPU film rather than seam failure, whereas Group 3 withstood the maximum air pressure of the testing machine without leakage.}

\subsection{Reusability}
\textcolor{black}{Reusability was primarily evaluated by repeatedly inflating and deflating 6 cm-diameter pneumatic actuators made of TPU film with petal-shaped 3D-printed structures (Fig.~\ref{appendix}.d), for 1,000 cycles under 60\% of the maximum air pressure determined in the airtightness experiment. Across five tested samples, none exhibited leakage after 1,000 cycles, and no 3D-printed structures detached. Only the TPU film showed slight wrinkling due to minor plastic deformation.}

\end{document}